\documentclass[aps,superscriptaddress,long,notitlepage
,balancelastpage,footinbib,prl, floatfix,
twocolumn]{revtex4-1}
\pdfoutput=1
\usepackage{amsmath,mathtools,amssymb,amsthm,amsxtra,overpic,bbm,epsfig,url,bm}
\usepackage{mathrsfs}
\usepackage{color}
\usepackage[dvipsnames]{xcolor}
\usepackage{float}
\usepackage{enumitem}
\usepackage{lmodern}
\usepackage{slashed}
\usepackage{multirow}
\usepackage{appendix}
\usepackage{tabularx}
\usepackage[T1]{fontenc}
\usepackage{graphicx}
\usepackage{slashed}
\usepackage{ragged2e}

\DeclareUnicodeCharacter{03BC}{\ensuremath{\mu}}
\DeclareUnicodeCharacter{2032}{\ensuremath{'}}
\definecolor{nicered}{rgb}{0.5,0.,0.}
\definecolor{nicegreen}{rgb}{0.,0.5,0.}
\definecolor{niceblue}{rgb}{0.,0.,0.5}
\definecolor{malagapurple}{cmyk}{0.431, 1, 0, 0.118}
\definecolor{malagagreen}{cmyk}{1, 0, 0.164, 0.141}
\usepackage{hyperref}
\hypersetup{%
	colorlinks=true,
	linkcolor=malagagreen,
	filecolor=malagagreen,      
	urlcolor=malagapurple,
	citecolor=malagapurple,
}

\newcommand{\prlsection}[2]{{\it\textbf{#1}{#2}}---}
\makeatletter
\newcommand*{\balancecolsandclearpage}{%
	\close@column@grid
	\cleardoublepage
	\twocolumngrid
}

\newcommand{\0}{$0\nu\beta\beta$}

\newcolumntype{Y}{>{\centering\arraybackslash}X}

\setlist{nolistsep}

\newcommand{\dbar}{\overline{d}}

\newcommand{\ebar}{\overline{e}}

\newcommand{\Qbar}{\overline{Q}}

\newcommand{\eps}{\epsilon}

\makeatother
\begin{document}

\title{
Importance of Loop Effects in Probing Lepton Number Violation 
}

\author{\bf Luk\'{a}\v{s} Gr\'{a}f}
\email[E-mail: ]{\textcolor{malagagreen}{lukas.graf@nikhef.nl}}
\affiliation{Nikhef, Theory Group, Science Park 105, 1098 XG Amsterdam, The Netherlands}
\affiliation{Institute of Particle and Nuclear Physics, Faculty of Mathematics and Physics, Charles University in Prague, V Hole\v{s}ovi\v{c}k\'ach 2, 180 00 Praha 8, Czech Republic}

\author{\bf Chandan Hati}
\email[E-mail: ]{\textcolor{malagagreen}{chandan@ific.uv.es}}
\affiliation{Instituto de F\'isica Corpuscular (IFIC), Universitat de Valencia-CSIC, E-46980 Valencia, Spain}

\author{\bf Ana Mart\'{i}n-Gal\'{a}n}
\email[E-mail: ]{\textcolor{malagagreen}{ana.martin@ific.uv.es}}
\affiliation{Instituto de F\'isica Corpuscular (IFIC), Universitat de Valencia-CSIC, E-46980 Valencia, Spain}

\author{\bf Oliver Scholer}
\email[E-mail: ]{\textcolor{malagagreen}{scholer@berkeley.edu}}
\affiliation{Max-Planck-Institut f{\"u}r Kernphysik, Saupfercheckweg 1, 69117 Heidelberg, Germany}
\affiliation{Department of Physics, University of California, Berkeley, CA 94720, USA}

\begin{abstract}
The discovery of the lepton number violation would be a smoking gun signal for physics beyond the Standard Model, and its most sensitive probe is the search for neutrinoless double beta decay ($0\nu\beta\beta$). Working in the framework of the Standard Model Effective Field Theory (SMEFT), we show that one-loop effects can remarkably improve the tree-level bounds on the new-physics scales for several dimension-7 operators. Using ultraviolet model examples, we then showcase the competition among $0\nu\beta\beta$ contributions induced by dimension-7 and loop-level dimension-5 SMEFT operators.
\end{abstract}

\maketitle

\prlsection{Introduction}{.}
Observation of Lepton Number Violation (LNV) would be a ground-breaking discovery and indisputable evidence for physics beyond the Standard Model (BSM). It can be intimately related to the key puzzles of contemporary particle physics, particularly to the nonzero neutrino masses indicated by the observed neutrino oscillations and the observed matter-antimatter asymmetry of the Universe. The celebrated seesaw mechanisms for Majorana neutrino masses~\cite{Minkowski:1977sc,Gell-Mann:1979vob,Yanagida:1979as,Mohapatra:1979ia,Schechter:1980gr,Ma:1998dn} (often extended in various forms to bring the new physics energy scales within the experimental reach) along with various radiative neutrino mass models~\cite{Cai:2017jrq}, have been explored in the literature to reproduce the tiny active neutrino masses, which is achieved by either a heavy-scale or a loop suppression.

From a model-independent viewpoint employing nonrenormalizable effective interactions respecting the local gauge symmetry of the SM, often referred to as the Standard Model Effective Field theory (SMEFT), the three types of conventional seesaw mechanisms correspond to the three minimal tree-level realizations for the nonrenormalizable lepton-number-violating dimension-5 Weinberg operator~\cite{Ma:1998dn}. Besides that, there exist a vast number of models realizing the neutrino masses radiatively, thus, giving rise to the Weinberg operator at a certain loop level~\cite{Bonnet:2012kz,AristizabalSierra:2014wal,Cepedello:2018rfh}. These scenarios can, however, lead at the same time to tree-level realization of the higher-dimensional LNV operators, which always appear at odd dimensions of SMEFT~\cite{Kobach:2016ami,Babu:2001ex,deGouvea:2007qla}. A priori, there is no reason for the LNV to necessarily manifest at the tree level at dimension 5; see, e.g.~Ref.~\cite{Fridell:2024pmw} for a study of various UV completions with the tree-level realization of dimension-7 operators where dimension-5 operators are realized at loop level. In such scenarios, the loop-induced lower-dimensional operators compete with the tree-level higher-dimensional operators, which may lead to important implications for the constraints on the LNV effective interactions obtained from the current experimental limits (in the bottom-up picture) and for making experimental predictions for low-energy LNV observable rates starting from simple, realistic UV models (in the top-down picture). 
A simplistic scale estimate for the contribution of LNV higher-dimensional operators to light neutrino masses by closing the loops of light SM particles~\cite{deGouvea:2007qla} suggests that the contribution to LNV observables like neutrinoless double beta decay (\0) from the light-neutrino-exchange mechanism realized at the loop level can potentially be more important compared to the direct contribution from the higher-dimensional operators~\footnote{This naive correlation between contribution to absolute light neutrino mass and the light-neutrino-exchange mechanism contribution to \0 (often parametrised by the effective neutrino mass parameter $m_{\beta\beta}$ at low-energies) can be broken e.g.~ if a cancellation can be arranged such that $m_{\beta\beta}$ is vanishingly small but the rate of \0 is not.}. However, such loop effects have been estimated using only the mixing contributions, mainly in the context of a simple naturalness-based correction to the absolute scale of neutrino masses, without a direct link to LNV experimental observables. 

In this letter, taking the example of one of the most promising experimental LNV observables, \0, we demonstrate for the first time using a proper SMEFT framework, including the full 
matching and mixing contributions, how the contributions induced at loop level can lead to large corrections changing the existing tree-level results in the literature in both the top-down and the bottom-up approaches. In particular, we focus on the inter-dimensional mixing among dimension-7 and -5 operators to highlight how loop-induced contributions to $0\nu\beta\beta$ can lead to significantly stronger constraints on many higher-dimensional LNV operators compared to existing ones. On the flip side, we also demonstrate that the observable rate for \0 can change by several orders of magnitude due to the interplay of loop-level matching and mixing effects. 

\prlsection{Loop effects for LNV Operators and its implications}{.}\label{sec:rges}
In the framework of SMEFT, the effective interactions beyond the renormalizable SM can be expressed in the form
\begin{align}
 	\label{eq:wilsonLNV}
 	\hspace{-1.5 mm}\mathcal{L}_\text{eff}= C^{(5)}_{LH}\mathcal{O}^{(5)}_{LH} + \hspace{-1.0 mm}  \sum_i C^{(6)}_i \mathcal{O}^{(6)}_i+ \hspace{-1.0 mm} \sum_j C_j^{(7)}\mathcal{O}_j^{(7)} + \dots \, ,
 \end{align}
 where $C^{(d)}$'s denote the dimensionful ($[M]^{4-d}$) Wilson coefficients and $\mathcal{O}^{(d)}$'s are the $d$-dimensional operators. The interactions violating the lepton number by two units appear only at the odd dimensions~\cite{Kobach:2016ami}. In this letter, we will primarily focus on LNV operators at dimension 7, the next-to-minimal dimensional LNV operators after the single dimension-5 Weinberg operator $\mathcal{O}^{(5)}_{LH} = \eps_{ij}\eps_{mn}(L_i^TCL_m )H_j H_n$.  We will follow the notation of Ref.~\cite{Cirigliano:2017djv} in denoting the 12 independent LNV dimension-7 operators, reiterated in the \textit{End Matter} for easy reference. 
 
 In the top-down picture, the loop level matching to different dimensional LNV operators needs to be done on an individual UV complete model basis after integrating out the heavy NP degrees of freedom at a given scale $\Lambda$, followed by taking into account any subsequent mixing between same and different dimensional operators to the scale of the experimental observable. On the other hand, when employing the bottom-up approach for deriving the constraints on the NP scale under a single SMEFT operator dominance assumption, operator mixing is of main relevance. Loop-level mixing effects among operators can be systematically analyzed using the renormalization group equations (RGEs)~\cite{Manohar:2018aog}. The running of the couplings in $d\leq4$ is given by the usual $\beta$-functions of the SM. Considering the LNV operators up to dimension 7, the RGEs that dictate the mixing for dimension-5 and dimension-7 operators can be expressed up to $\mathcal{O}(\Lambda^{-3})$ as~\cite{Zhang:2023kvw}
 \begin{eqnarray}\label{eq:LNV_rge1}
	\hspace{-4 mm}
 {\dot{C}_{}^{(5)}} \hspace{-2 mm}&=&  \gamma^{(5,5)} C_{}^{(5)} \hspace{-1 mm}+ \hat{\gamma}^{(5,5)} C_{}^{(5)} C_{}^{(5)}  C_{}^{(5)} \hspace{-1 mm}+ \gamma^{(5,6)}_i C_{}^{(5)}  C_{i}^{(6)}\nonumber\\ && + \gamma^{(5,7)}_i C_{i}^{(7)} ,
	\\
	\hspace{-4 mm}
 {\dot{C}_{i}^{(7)}} \hspace{-2 mm} &=&  \gamma^{(7,7)}_{ij} C_{j}^{(7)} \hspace{-1 mm}+ \gamma^{(7,5)}_i C_{}^{(5)}  C_{}^{(5)}  C_{}^{(5)} \hspace{-1 mm} + \gamma^{(7,6)}_{ij} C_{}^{(5)}  C_{j}^{(6)} , \label{eq:LNV_rge2}
\end{eqnarray}
where dots denote the rate of change w.r.t.~ the logarithm of the scale, while $\gamma$'s denote the anomalous dimension tensors (or matrices). The repeated indices are summed over, and the flavor indices are suppressed for brevity. The RGEs for dimension-6 operators can be found in~\cite{Jenkins:2013zja,Jenkins:2013wua,Alonso:2013hga,Alonso:2014zka}. RGEs corresponding to LNV operators have been partially derived and used in the context of neutrino masses in~\cite{Babu:1993qv,Chankowski:1993tx,Antusch:2001ck,Chala:2021juk}, while in~\cite{Liao:2019tep,Zhang:2023kvw,Zhang:2023ndw} a complete set of anomalous dimension tensor entries have been recently derived~\footnote{In Ref.~\cite{Zhang:2023kvw}, it was shown that $\hat{\gamma}^{(5,5)}$ vanishes for the dimension-5 operator, while $\gamma^{(7,5)}$ vanishes for all dimension-7 operators except one, $\mathcal{O}^{(7)}_{LH}$.}. The mixing among dimension-7 operators themselves ($\gamma^{(7,7)}$) has been known to provide important effects in the context of light neutrino masses originating from LNV effective interaction~\cite{Cirigliano:2017djv,Chala:2021juk}. In what follows, we will show, for the first time, how the loop-level operator matching and operator mixing can dominate over the tree-level matching for the LNV observables, such as $0\nu\beta \beta$.
\begin{figure}[h]
    \centering
    \includegraphics[width=1\columnwidth]{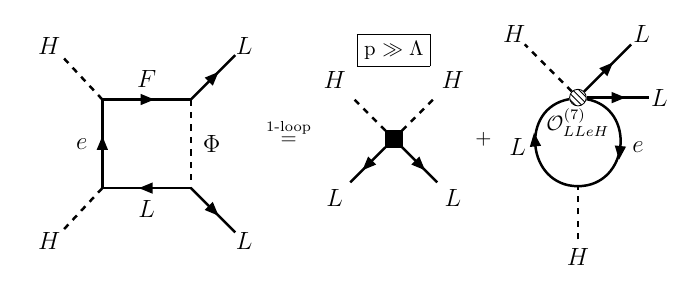}
    \caption{One-loop matching and mixing contributions to the dimension-5 Weinberg operator (absent at tree level) in a UV-model example generating only $\mathcal{O}^{(7)}_{LL\ebar H}$ at tree level.}
    \label{fig:mix_dig}
\end{figure}

In many UV constructions, the LNV can manifest directly at a higher dimension, e.g.~ at dimension 7 without any tree-level contribution to $\mathcal{O}^{(5)}_{LH}$~\cite{Fridell:2024pmw}. In such case, even though absent at the tree level, $\mathcal{O}^{(5)}_{LH}$ is induced via operator mixing at loop level; see, e.g.~ Fig.~\ref{fig:mix_dig}. Given the quite different sensitivities of LNV observables, such as $0\nu\beta\beta$, to dimension-5 and dimension-7 operators, the loop-induced contributions can become competitive and even dominant over the direct tree-level contributions from the original operator. Hence, they can provide constraints orders of magnitude stronger than those obtained using only the tree-level matching~\cite{Graesser:2016bpz,Cirigliano:2017djv,Fridell:2023rtr}. In certain cases, the LNV SMEFT operators do not contribute to $0\nu\beta\beta$ at the tree level, and the current best constraints available in the literature stem from other probes (e.g.~ from nonstandard muon decay for the case of the leptonic operator $\mathcal{O}^{(7)}_{LL\ebar H}$). We will see that even in such cases, the loop-level operator mixing can lead to a $0\nu\beta\beta$ bound more stringent than any other experimental limit. 

\prlsection{Loop-induced constraints on LNV scale from $0\nu\beta\beta$}{.}\label{sec:eftlimits}
It is well-known that \0 provides the most stringent constraints on LNV in the electron flavour. In fact, it is sensitive to some of the dimension-7 and dimension-9 LNV SMEFT operators for NP scales well beyond the current and future collider reach~\cite{Scholer:2023bnn}. The half-life for \0 can be expressed as~\cite{Cirigliano:2017djv, Cirigliano:2018yza}
\begin{align}
    T^{-1}_{1/2} &= g_A^4\sum_k G_{0k} |\mathcal{A}_k(\{C_i\})|^2,
\end{align}
where $g_A \simeq 1.27$ is the axial coupling, $G_{0k}$ denotes the phase space factors (PSFs)~\cite{Kotila:2012zza, Stefanik:2015twa}, while most of the physics is captured by the so-called sub-amplitudes $\mathcal{A}_k(\{C_i\})$, which depend on nuclear matrix elements (NMEs)~\cite{Deppisch:2020ztt}, chiral-EFT low energy constants (LECs)~\cite{Gasser:1983yg, Weinberg:1990rz, Weinberg:1991um, Scherer:2002tk, Scherer:2005ri, Bhattacharya:2016zcn, Nicholson:2018mwc, Cirigliano:2020dmx, Cirigliano:2021qko, Wirth:2021pij}, and the Wilson coefficients of lepton-number-violating higher-dimensional operators labelled by $\{C_i\}$. The full expressions of the sub-amplitudes are rather lengthy and can be found, for instance, in Refs.~\cite{Cirigliano:2017djv, Cirigliano:2018yza}. For the nuclear matrix elements, we employ numerical values from Ref.~\cite{Deppisch:2020ztt}, and for the PSFs, we use the values obtained from the exact solutions to the electron radial wave functions in a point-like nucleus approximation~\cite{Stefanik:2015twa, Scholer:2023bnn}. 

\begin{figure}[t]
    \centering
    \includegraphics[width=1\linewidth]{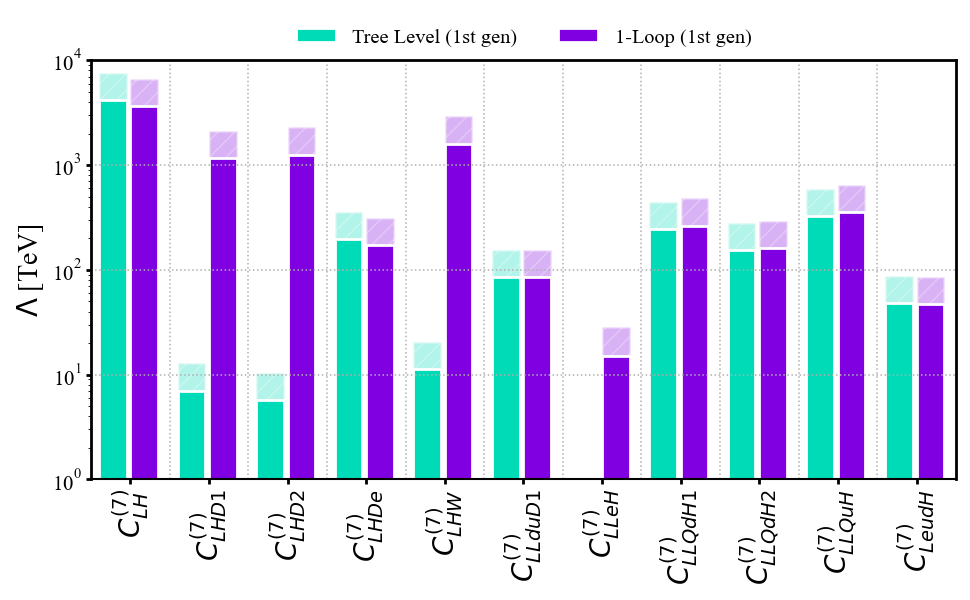}
    \caption{ Comparison of the \0 sensitivities to the NP scale $(\Lambda)$ of different LNV dimension-7 SMEFT operators between only tree-level matching (green) and with one-loop mixing effects included (purple) between $\Lambda$ and the electroweak scale, under the assumption of a single SMEFT operator dominance. The dark and light parts of the bands correspond to the sensitivities from the complete KamLAND-Zen results~\cite{KamLAND-Zen:2024eml} and the projected sensitivity of the next-generation ton-scale experiment nEXO~\cite{nEXO:2021ujk}, respectively.
    }
    \label{fig:limits_0nubb}
\end{figure}

The limits on the dimension-7 SMEFT Wilson coefficients $C^{(7)}_i$ can be translated into bounds on the corresponding new-physics scale $\Lambda$ as $\Lambda^3 \simeq 1/{C_i^{(7)}}$. The most up-to-date tree-level constraints imposed by \0 searches can be found, for instance, in Ref.~\cite{Fridell:2023rtr}, which updated the constraints in Refs.~\cite{Cirigliano:2017djv, Cirigliano:2018yza}, taking into account the latest experimental results from KamLAND-Zen~\cite{KamLAND-Zen:2022tow}. We find that these tree-level constraints change significantly for several of the dimension-7 SMEFT operators once the mixing among them and the dimension-5 Weinberg operator is taken into account. In Fig.~\ref{fig:limits_0nubb}, we show the sensitivity of \0 in current and future generations of experiments to the various dimension-7 SMEFT operators without and with one-loop mixing. We find that the operators $\mathcal{O}_{LHD1}^{(7)}$, $\mathcal{O}_{LHD2}^{(7)}$ and $\mathcal{O}_{LHW}^{(7)}$ experience a striking increase in sensitivity. The one-loop level limits are stronger by more than two orders of magnitude when compared with the tree-level results. For these operators, both the loop-level mixing among different dimension-7 operators themselves, as well as the loop mixing with the dimension-5 Weinberg operator, lead to a fairly large contribution to \0 compared to the simple tree-level contribution. What further sticks out is the purely leptonic operator $\mathcal{O}_{LL\bar{e}H}^{(7)}$, which does not trigger \0 at the tree level. However, at the one-loop level, \0 also yields a limit of tens of TeV for the NP scale associated with this operator, which stems from its mixing with the Weinberg operator. Intriguingly, the loop-level \0 constraint surpasses the current best limit on this operator, imposed by LNV muon decay~\cite{Cirigliano:2017djv,Fridell:2023rtr}, by two orders of magnitude. Less striking but significant changes are also observed for $\mathcal{O}_{LH}^{(7)}$ and $\mathcal{O}_{LLduD1}^{(7)}$. 


\prlsection{Impact of loop mixing for UV complete theory predictions}{.}\label{sec:uvbounds}
From the newly obtained more stringent limits on several of the dimension-7 SMEFT operators above, it must be apparent that, at least for these operators, the prediction of \0 rates should see large increments. However, several interesting features and subtleties become apparent when the one-loop effects are included in matching a UV complete model to evaluate \0 rates using the top-down EFT approach. To highlight these, let us consider three specific examples, the last two of which have been previously studied in the literature in other contexts.

\begin{figure*}[t]
    \centering
    \includegraphics[width=1\linewidth]{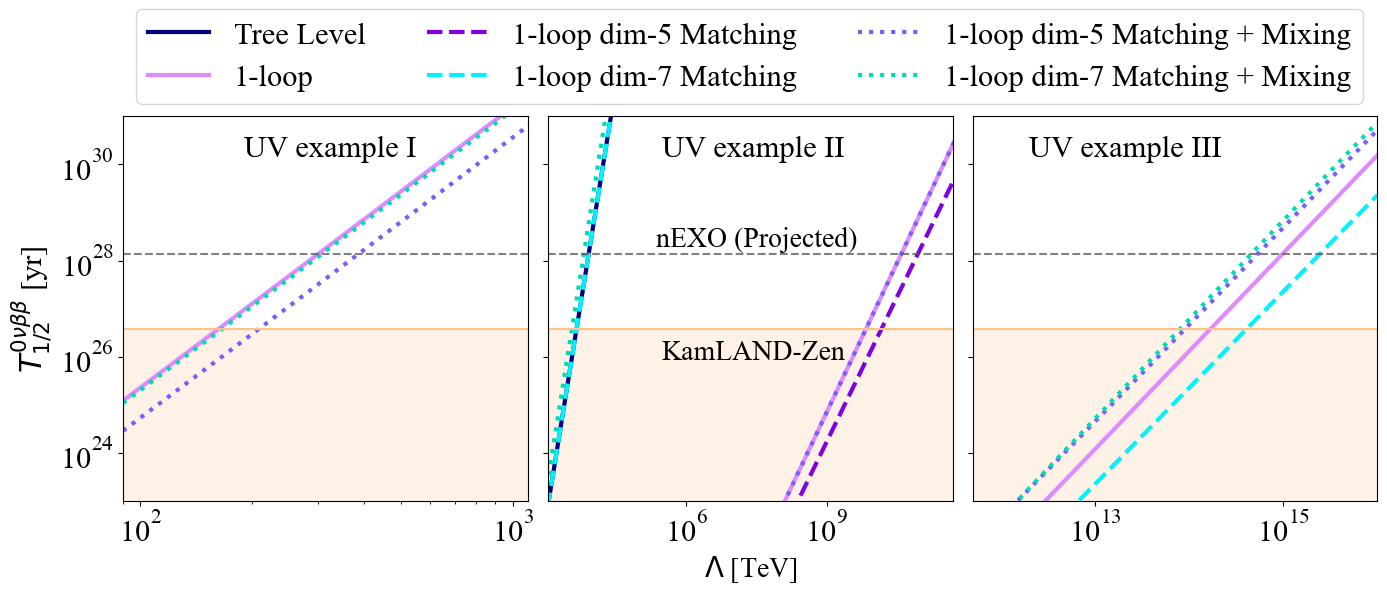}
    \caption{ The \0 half-life due to tree-level and different one-loop level contributions to LNV dimension-5 and dimension-7 SMEFT operators for the three UV-complete models described in the text. Each point on the marked lines corresponds to the predicted final half-life if the heavy NP fields with mass $M_{\text {NP}}=\Lambda$ are integrated out at the scale $\Lambda$ with only the NP couplings to first generation assumed to be nonvanishing and unity (except $3\times 3$ antisymmetric coupling $f_1$ in Eq.~\ref{model_1_lag}, where all non-diagonal entries are assumed to be unity in magnitude). The orange shading and the black-dashed lines show the currently best KamLAND-Zen sensitivity~\cite{KamLAND-Zen:2024eml} and the projected sensitivity of nEXO~\cite{nEXO:2021ujk}, respectively. The solid pink curve represents the final total one-loop contribution, encompassing both dimension-5 and dimension-7 contributions. The dashed curves are just matching contributions, ignoring any loop mixing effects due to the RGEs. The dotted lines include effects due to both matching and RGE running.} 
    \label{fig:model_top-down}
\end{figure*}
\paragraph{UV example I:} The first UV model example we will consider is the SM extended by a vector-like fermion $F:\{1,2,3/2\}$ and a singlet charged scalar $\Phi:\{1,1,1\}$, where we denote the charges in notation $\{SU(3)_c$, $SU(2)$, $U(1)_Y\}$. Here and in the following discussions, we use the convention such that electric charge $Q=T_3+Y$ and for simplicity we also assume all heavy new physics fields to be of approximately the same mass $M_\mathrm{NP}\sim\Lambda$~\footnote{If the heavy degrees of freedom are hierarchical in mass, then one should employ step-by-step matching and mixing integerating out one heavy degree of freedom at a time starting from the UV model. In such a scenario, one would expect logarithmic corrections due to the running between the two hierarchical heavier scales.}.
The relevant new interactions are given by 
\begin{align}\label{model_1_lag}
  \mathcal{L}_{I} \supset f_1 \overline{L}  i\tau_2 {L}^C \Phi^{\dagger} + f_2 \overline{e^C}  F H^{\dagger} + f_3\overline{F} i\tau_2 {L}^C \Phi\, ,
\end{align}
where $L$ denotes the SM lepton doublet, $H$ denotes the SM Higgs doublet, $\tau_2$ denotes the second Pauli matrix, and hermitian conjugated terms are implied. At tree-level, this model only generates the dimension-7 operator $\mathcal{O}_{LL\bar{e}H}^{(7)}$ without any other dimension-7 or -5 contributions; hence, there is no tree-level \0 contribution. At the one-loop level, this model can generate a matching contribution to the Weinberg operator at the new physics scale $\mu=\Lambda$ (when $F$ and $\Phi$ are integrated out), depicted by the middle diagram of Fig.~\ref{fig:mix_dig}~\footnote{For the figure we make the benchmark choice of fixing the NP couplings to unity. However, for smaller values of the NP couplings, both dimension-5 and -7 lines shift upwards, with slightly different slopes because of a different $\Lambda$ dependence.}. The one-loop matching contribution has the form
\begin{align}
\Delta C_5^{\alpha \beta} =& \frac{M_F F_{\text{AS}}^{\alpha \beta} (y_l,f_1^\dagger,f_2,f_3)}{32 \pi^2 (M_F^2 - M_{\Phi}^2)} \ln \left( \frac{M_F^2}{M_\Phi^2} \right) \, ,
\end{align}
where $y_l$ is the SM lepton Yukawa coupling and the function $F_{\text{AS}}^{\alpha\beta}\equiv
y_l^{\gamma \lambda} [(f_1^{\dagger\gamma \alpha}  -  f_1^{\dagger \alpha \gamma}) f_2^{ \lambda}  f_3^{\beta} 
 + \alpha\leftrightarrow \beta]$ is vanishing for a symmetric $f_1$ Yukawa matrix. The mixing of operators between the NP scale $\Lambda$ and the electroweak scale also generates contributions to the dimension-5 Weinberg operator and, at a sub-dominant level, to other dimension-7 operators. The dimension-5 mixing contributions correspond to the one-loop renormalization correction arising from closing the two of the legs of $\mathcal{O}_{LL\bar{e}H}^{(7)}$ via the SM lepton Yukawa coupling, as shown in the right diagram of Fig.~\ref{fig:mix_dig}. Finally, the left-most panel of Fig.~\ref{fig:model_top-down} shows the different contributions at the one-loop level arising in this model assuming that the NP only couples to electrons, demonstrating that the mixing contribution to the Weinberg operator dominates the contributions to \0~\footnote{The behaviour of the different contributions and the conclusion remains very similar if we instead assume a flavour universal structure for $f_1$ Yukawa matrix, with slight shift in scale.}.  The high sensitivity of \0 to the Weinberg operator allows for testing this kind of scenario, giving orders of magnitude better sensitivity compared to the currently known best probe for such a scenario in the form of LNV muon decay~\cite{Cirigliano:2017djv,Fridell:2023rtr}.

\paragraph{UV example II:} The second UV model example we will consider is the SM extended by a vector-like fermion $\Sigma:\{1,3,1\}$ and a scalar $S:\{1,4,3/2\}$. The relevant new interactions are given by
\begin{align}
  \mathcal{L}_{II}\supset h_1 \overline{L^C} \Sigma H^{\dagger} + h_2 \overline{\Sigma} L S + \lambda_S H H H S^\dagger \, .
\end{align}
This model, proposed in Ref.~\cite{Babu:2009aq}, is one of the earliest known model examples where the dimension-7 operator $\mathcal{O}_{LH}^{(7)}$ is known to be generated at tree level without any tree-level Weinberg operator. The dimension-5 contribution is only generated at one-loop order when two of the Higgs legs of $\mathcal{O}_{LH}^{(7)}$ are closed into a loop using the quadratic Higgs term. The relevant matching contribution to the Weinberg operator has the form
\begin{align}
\Delta C_5^{\alpha \beta}=\frac{M_\Sigma}{32\pi^2} \frac{\lambda_S^{\dagger} (h_1^{\alpha} h_2^{\beta} + h_1^{\beta} h_2^{\alpha} )}{(M_\Sigma^2 - M_{S}^2)} \ln \left( \frac{M_S^2}{M_\Sigma^2} \right) \, ,
\end{align}
which dominates the \0 rate over the tree-level contributions from the dimension-7 operator $\mathcal{O}_{LH}^{(7)}$ within the whole range of energies accessible at the current and future generation experiments, as shown in the middle panel of Fig.~\ref{fig:model_top-down}. This is, again, due to the vastly different sensitivity to the \0 rate triggered by the dimension-5 Weinberg operator compared to the one induced by the dimension-7 operators.
%
\paragraph{UV example III:} The third and final UV model example we will consider is the extension of the SM gauge group by a global $U'(1)$ symmetry and the field content with a vector-like triplet fermion $\Sigma': \{1,3,1,q\}$, and two doublet scalars $\chi :\{1,2,3/2,q\}$, and $\eta :\{1,2,1/2,q\}$, where the forth quantum number in the parentheses denote the $U'(1)$ charge. The relevant new interactions are given by
\begin{align}
   \mathcal{L}_{III} \supset y_1 \overline{L} \Sigma' \tilde{\chi} + y_2 \overline{\hat{L}^C} \Sigma' \tilde{\eta} +\kappa (\chi^\dagger H)(\tilde{\eta}^\dagger H) \, 
 \end{align}
where $\hat{L}=i\tau_2 L$ and $\tilde{\chi}=i\tau_2 \chi^*$. This model was proposed in Ref.~\cite{Aoki:2020til} as a UV completion of the $\mathcal{O}_{LHD1}^{(7)}$ operator at one-loop order \footnote{Note that we have slightly modified some of the interactions to make the quantum number assignments consistent with the interactions, which seemed to contain typos in the original reference.}. Because of the $U'(1)$ global symmetry, the Weinberg operator is not generated at the tree level; however, it can originate at one-loop order, i.e.~ the same order as the $\mathcal{O}_{LHD1}^{(7)}$ operator \footnote{In the absence of any additional symmetry, the tree-level UV completions of $\mathcal{O}_{LHD1}^{(7)}$ involve one of the three particles leading to tree-level dimension-5 seesaw realizations. Same is also true for $\mathcal{O}_{LHD2}^{(7)}$, $\mathcal{O}_{LHW}^{(7)}$ and $\mathcal{O}_{LHB}^{(7)}$.}. Following a careful examination we find that this model, in fact, also generates one-loop matching contributions to $\mathcal{O}_{LH}^{(7)}$, $\mathcal{O}_{LHW}^{(7)}$, $\mathcal{O}_{LHB}^{(7)}$, $\mathcal{O}_{LHD2}^{(7)}$, $\mathcal{O}_{LL\bar{e} H}^{(7)}$, $\mathcal{O}_{LLQ\bar{d} H1}^{(7)}$, $\mathcal{O}_{LLQ\bar{d} H2}^{(7)}$ and $\mathcal{O}_{LL\bar{Q}uH}^{(7)}$
in our physical basis, leading to a similar contribution to \0
as the dimension-5 Weinberg operator, which can be seen in the right panel of Fig.~\ref{fig:model_top-down}. The individual contributions of the different dimension-7 Wilson coefficients are included in the End matter. Therefore, this model provides a counter-example to the previous two scenarios, in which the loop-level dimension-5 dominated over the dimension-7 contributions~\footnote{Interestingly, $\mathcal{O}_{LH}^{(7)}$ (which can directly contribute to light neutrino masses after the electroweak symmetry breaking) provides the most dominant contribution among all the dimension-7 operators for $\Lambda>4\times10^{12}$.}.

\prlsection{Discussion}{.}\label{sec:discussion}
In this work, we have primarily focused on \0 to demonstrate the loop mixing effects for its prominent sensitivity to LNV new physics coupling to the first-generation leptons. However, similar effects are also expected for other observables like the LNV meson decays and collider searches. For the off-resonance collider searches of effective contact interactions, the current best limits are relatively poor~\cite{Fridell:2023rtr,Aoki:2020til,Fuks:2020zbm} for the mixing effects between heavy NP scale and collider energy scale (e.g.~ center-of-mass energy) to be of immediate importance. On the other hand, for probes like the $K\rightarrow \pi \nu\nu$ decays, these effects can be relevant and lead to new nontrivial interplay between the dimension-5 and dimension-7 LNV operators induced by the mixing. 
 
The effects discussed in this work imply that a careful implementation of the operator mixing triggers important contributions to the standard light-neutrino-exchange mechanism, hinting at the challenging nature of the potential efforts to distinguish exotic high-energy \0 mechanism from observations of this rare decay itself~\cite{Deppisch:2006hb,Graf:2022lhj}. Also, assuming a suppressed Weinberg-operator tree-level contribution to \0, the herein-discussed loop-induced contributions induced by the higher-dimensional LNV operators may often be of greater importance than the ambiguities stemming from the nuclear structure computations.
 
 In conclusion, we have demonstrated how the loop-induced contributions to \0 rate can often entirely dominate over higher-dimensional tree-level rates and, hence, improve the constraints on the NP scale by several orders of magnitude. For realistic UV complete model examples, these effects can enhance the experimental predictions for \0 rates by several orders of magnitude compared to purely tree-level calculations. As such, these corrections are of significant relevance for phenomenology, and they nicely complement the mosaic of recent improvements in the field of \0.


\begin{acknowledgments}
\prlsection{Acknowledgments}{.}
The authors are thankful to  Frank F. Deppisch, Jordy de Vries and Martin K. Hirsch for carefully reading the manuscript and for providing valuable comments. The authors acknowledge Javier Fuentes-Mart\'{i}n, Jose Santiago and Di Zhang for some very useful correspondences. L.~G.~acknowledges support from the Dutch Research Council (NWO) under project number VI.Veni.222.318, and from Charles University through project PRIMUS/24/SCI/013. 
C.~H. is funded by the Generalitat Valenciana under Plan Gen-T via CDEIGENT grant No. CIDEIG/2022/16. A.~M. is funded by Generalitat Valenciana via grant No. CIDEXG/2022/20.  C.~H. and A.~M.~ also acknowledge support from the Spanish grants PID2023-147306NB-I00 and CEX2023-001292-S (MCIU/AEI/10.13039/501100011033). O.~S. acknowledges support by the Alexander von Humboldt Foundation under the Feodor Lynen Research Fellowship program and by the National Science Foundation under cooperative agreement 2020275.
\end{acknowledgments}
\bibliographystyle{apsrev4-1}
\twocolumngrid
\bibliography{references.bib}

\begin{thebibliography}{62}%
\makeatletter
\providecommand \@ifxundefined [1]{%
 \@ifx{#1\undefined}
}%
\providecommand \@ifnum [1]{%
 \ifnum #1\expandafter \@firstoftwo
 \else \expandafter \@secondoftwo
 \fi
}%
\providecommand \@ifx [1]{%
 \ifx #1\expandafter \@firstoftwo
 \else \expandafter \@secondoftwo
 \fi
}%
\providecommand \natexlab [1]{#1}%
\providecommand \enquote  [1]{``#1''}%
\providecommand \bibnamefont  [1]{#1}%
\providecommand \bibfnamefont [1]{#1}%
\providecommand \citenamefont [1]{#1}%
\providecommand \href@noop [0]{\@secondoftwo}%
\providecommand \href [0]{\begingroup \@sanitize@url \@href}%
\providecommand \@href[1]{\@@startlink{#1}\@@href}%
\providecommand \@@href[1]{\endgroup#1\@@endlink}%
\providecommand \@sanitize@url [0]{\catcode `\\12\catcode `\$12\catcode
  `\&12\catcode `\#12\catcode `\^12\catcode `\_12\catcode `\%12\relax}%
\providecommand \@@startlink[1]{}%
\providecommand \@@endlink[0]{}%
\providecommand \url  [0]{\begingroup\@sanitize@url \@url }%
\providecommand \@url [1]{\endgroup\@href {#1}{\urlprefix }}%
\providecommand \urlprefix  [0]{URL }%
\providecommand \Eprint [0]{\href }%
\providecommand \doibase [0]{http://dx.doi.org/}%
\providecommand \selectlanguage [0]{\@gobble}%
\providecommand \bibinfo  [0]{\@secondoftwo}%
\providecommand \bibfield  [0]{\@secondoftwo}%
\providecommand \translation [1]{[#1]}%
\providecommand \BibitemOpen [0]{}%
\providecommand \bibitemStop [0]{}%
\providecommand \bibitemNoStop [0]{.\EOS\space}%
\providecommand \EOS [0]{\spacefactor3000\relax}%
\providecommand \BibitemShut  [1]{\csname bibitem#1\endcsname}%
\let\auto@bib@innerbib\@empty
\bibitem [{\citenamefont {Minkowski}(1977)}]{Minkowski:1977sc}%
  \BibitemOpen
  \bibfield  {author} {\bibinfo {author} {\bibfnamefont {P.}~\bibnamefont
  {Minkowski}},\ }\href {\doibase 10.1016/0370-2693(77)90435-X} {\bibfield
  {journal} {\bibinfo  {journal} {Phys. Lett. B}\ }\textbf {\bibinfo {volume}
  {67}},\ \bibinfo {pages} {421} (\bibinfo {year} {1977})}\BibitemShut
  {NoStop}%
\bibitem [{\citenamefont {Gell-Mann}\ \emph {et~al.}(1979)\citenamefont
  {Gell-Mann}, \citenamefont {Ramond},\ and\ \citenamefont
  {Slansky}}]{Gell-Mann:1979vob}%
  \BibitemOpen
  \bibfield  {author} {\bibinfo {author} {\bibfnamefont {M.}~\bibnamefont
  {Gell-Mann}}, \bibinfo {author} {\bibfnamefont {P.}~\bibnamefont {Ramond}}, \
  and\ \bibinfo {author} {\bibfnamefont {R.}~\bibnamefont {Slansky}},\
  }\href@noop {} {\bibfield  {journal} {\bibinfo  {journal} {Conf. Proc. C}\
  }\textbf {\bibinfo {volume} {790927}},\ \bibinfo {pages} {315} (\bibinfo
  {year} {1979})},\ \Eprint {http://arxiv.org/abs/1306.4669} {arXiv:1306.4669
  [hep-th]} \BibitemShut {NoStop}%
\bibitem [{\citenamefont {Yanagida}(1979)}]{Yanagida:1979as}%
  \BibitemOpen
  \bibfield  {author} {\bibinfo {author} {\bibfnamefont {T.}~\bibnamefont
  {Yanagida}},\ }\href@noop {} {\bibfield  {journal} {\bibinfo  {journal}
  {Conf. Proc. C}\ }\textbf {\bibinfo {volume} {7902131}},\ \bibinfo {pages}
  {95} (\bibinfo {year} {1979})}\BibitemShut {NoStop}%
\bibitem [{\citenamefont {Mohapatra}\ and\ \citenamefont
  {Senjanovic}(1980)}]{Mohapatra:1979ia}%
  \BibitemOpen
  \bibfield  {author} {\bibinfo {author} {\bibfnamefont {R.~N.}\ \bibnamefont
  {Mohapatra}}\ and\ \bibinfo {author} {\bibfnamefont {G.}~\bibnamefont
  {Senjanovic}},\ }\href {\doibase 10.1103/PhysRevLett.44.912} {\bibfield
  {journal} {\bibinfo  {journal} {Phys. Rev. Lett.}\ }\textbf {\bibinfo
  {volume} {44}},\ \bibinfo {pages} {912} (\bibinfo {year} {1980})}\BibitemShut
  {NoStop}%
\bibitem [{\citenamefont {Schechter}\ and\ \citenamefont
  {Valle}(1980)}]{Schechter:1980gr}%
  \BibitemOpen
  \bibfield  {author} {\bibinfo {author} {\bibfnamefont {J.}~\bibnamefont
  {Schechter}}\ and\ \bibinfo {author} {\bibfnamefont {J.~W.~F.}\ \bibnamefont
  {Valle}},\ }\href {\doibase 10.1103/PhysRevD.22.2227} {\bibfield  {journal}
  {\bibinfo  {journal} {Phys. Rev. D}\ }\textbf {\bibinfo {volume} {22}},\
  \bibinfo {pages} {2227} (\bibinfo {year} {1980})}\BibitemShut {NoStop}%
\bibitem [{\citenamefont {Ma}(1998)}]{Ma:1998dn}%
  \BibitemOpen
  \bibfield  {author} {\bibinfo {author} {\bibfnamefont {E.}~\bibnamefont
  {Ma}},\ }\href {\doibase 10.1103/PhysRevLett.81.1171} {\bibfield  {journal}
  {\bibinfo  {journal} {Phys. Rev. Lett.}\ }\textbf {\bibinfo {volume} {81}},\
  \bibinfo {pages} {1171} (\bibinfo {year} {1998})},\ \Eprint
  {http://arxiv.org/abs/hep-ph/9805219} {arXiv:hep-ph/9805219} \BibitemShut
  {NoStop}%
\bibitem [{\citenamefont {Cai}\ \emph {et~al.}(2017)\citenamefont {Cai},
  \citenamefont {Herrero-Garc\'\i{}a}, \citenamefont {Schmidt}, \citenamefont
  {Vicente},\ and\ \citenamefont {Volkas}}]{Cai:2017jrq}%
  \BibitemOpen
  \bibfield  {author} {\bibinfo {author} {\bibfnamefont {Y.}~\bibnamefont
  {Cai}}, \bibinfo {author} {\bibfnamefont {J.}~\bibnamefont
  {Herrero-Garc\'\i{}a}}, \bibinfo {author} {\bibfnamefont {M.~A.}\
  \bibnamefont {Schmidt}}, \bibinfo {author} {\bibfnamefont {A.}~\bibnamefont
  {Vicente}}, \ and\ \bibinfo {author} {\bibfnamefont {R.~R.}\ \bibnamefont
  {Volkas}},\ }\href {\doibase 10.3389/fphy.2017.00063} {\bibfield  {journal}
  {\bibinfo  {journal} {Front. in Phys.}\ }\textbf {\bibinfo {volume} {5}},\
  \bibinfo {pages} {63} (\bibinfo {year} {2017})},\ \Eprint
  {http://arxiv.org/abs/1706.08524} {arXiv:1706.08524 [hep-ph]} \BibitemShut
  {NoStop}%
\bibitem [{\citenamefont {Bonnet}\ \emph {et~al.}(2012)\citenamefont {Bonnet},
  \citenamefont {Hirsch}, \citenamefont {Ota},\ and\ \citenamefont
  {Winter}}]{Bonnet:2012kz}%
  \BibitemOpen
  \bibfield  {author} {\bibinfo {author} {\bibfnamefont {F.}~\bibnamefont
  {Bonnet}}, \bibinfo {author} {\bibfnamefont {M.}~\bibnamefont {Hirsch}},
  \bibinfo {author} {\bibfnamefont {T.}~\bibnamefont {Ota}}, \ and\ \bibinfo
  {author} {\bibfnamefont {W.}~\bibnamefont {Winter}},\ }\href {\doibase
  10.1007/JHEP07(2012)153} {\bibfield  {journal} {\bibinfo  {journal} {JHEP}\
  }\textbf {\bibinfo {volume} {07}},\ \bibinfo {pages} {153} (\bibinfo {year}
  {2012})},\ \Eprint {http://arxiv.org/abs/1204.5862} {arXiv:1204.5862
  [hep-ph]} \BibitemShut {NoStop}%
\bibitem [{\citenamefont {Aristizabal~Sierra}\ \emph
  {et~al.}(2015)\citenamefont {Aristizabal~Sierra}, \citenamefont {Degee},
  \citenamefont {Dorame},\ and\ \citenamefont
  {Hirsch}}]{AristizabalSierra:2014wal}%
  \BibitemOpen
  \bibfield  {author} {\bibinfo {author} {\bibfnamefont {D.}~\bibnamefont
  {Aristizabal~Sierra}}, \bibinfo {author} {\bibfnamefont {A.}~\bibnamefont
  {Degee}}, \bibinfo {author} {\bibfnamefont {L.}~\bibnamefont {Dorame}}, \
  and\ \bibinfo {author} {\bibfnamefont {M.}~\bibnamefont {Hirsch}},\ }\href
  {\doibase 10.1007/JHEP03(2015)040} {\bibfield  {journal} {\bibinfo  {journal}
  {JHEP}\ }\textbf {\bibinfo {volume} {03}},\ \bibinfo {pages} {040} (\bibinfo
  {year} {2015})},\ \Eprint {http://arxiv.org/abs/1411.7038} {arXiv:1411.7038
  [hep-ph]} \BibitemShut {NoStop}%
\bibitem [{\citenamefont {Cepedello}\ \emph {et~al.}(2018)\citenamefont
  {Cepedello}, \citenamefont {Fonseca},\ and\ \citenamefont
  {Hirsch}}]{Cepedello:2018rfh}%
  \BibitemOpen
  \bibfield  {author} {\bibinfo {author} {\bibfnamefont {R.}~\bibnamefont
  {Cepedello}}, \bibinfo {author} {\bibfnamefont {R.~M.}\ \bibnamefont
  {Fonseca}}, \ and\ \bibinfo {author} {\bibfnamefont {M.}~\bibnamefont
  {Hirsch}},\ }\href {\doibase 10.1007/JHEP10(2018)197} {\bibfield  {journal}
  {\bibinfo  {journal} {JHEP}\ }\textbf {\bibinfo {volume} {10}},\ \bibinfo
  {pages} {197} (\bibinfo {year} {2018})},\ \bibinfo {note} {[Erratum: JHEP 06,
  034 (2019)]},\ \Eprint {http://arxiv.org/abs/1807.00629} {arXiv:1807.00629
  [hep-ph]} \BibitemShut {NoStop}%
\bibitem [{\citenamefont {Kobach}(2016)}]{Kobach:2016ami}%
  \BibitemOpen
  \bibfield  {author} {\bibinfo {author} {\bibfnamefont {A.}~\bibnamefont
  {Kobach}},\ }\href {\doibase 10.1016/j.physletb.2016.05.050} {\bibfield
  {journal} {\bibinfo  {journal} {Phys. Lett. B}\ }\textbf {\bibinfo {volume}
  {758}},\ \bibinfo {pages} {455} (\bibinfo {year} {2016})},\ \Eprint
  {http://arxiv.org/abs/1604.05726} {arXiv:1604.05726 [hep-ph]} \BibitemShut
  {NoStop}%
\bibitem [{\citenamefont {Babu}\ and\ \citenamefont
  {Leung}(2001)}]{Babu:2001ex}%
  \BibitemOpen
  \bibfield  {author} {\bibinfo {author} {\bibfnamefont {K.~S.}\ \bibnamefont
  {Babu}}\ and\ \bibinfo {author} {\bibfnamefont {C.~N.}\ \bibnamefont
  {Leung}},\ }\href {\doibase 10.1016/S0550-3213(01)00504-1} {\bibfield
  {journal} {\bibinfo  {journal} {Nucl. Phys. B}\ }\textbf {\bibinfo {volume}
  {619}},\ \bibinfo {pages} {667} (\bibinfo {year} {2001})},\ \Eprint
  {http://arxiv.org/abs/hep-ph/0106054} {arXiv:hep-ph/0106054} \BibitemShut
  {NoStop}%
\bibitem [{\citenamefont {de~Gouvea}\ and\ \citenamefont
  {Jenkins}(2008)}]{deGouvea:2007qla}%
  \BibitemOpen
  \bibfield  {author} {\bibinfo {author} {\bibfnamefont {A.}~\bibnamefont
  {de~Gouvea}}\ and\ \bibinfo {author} {\bibfnamefont {J.}~\bibnamefont
  {Jenkins}},\ }\href {\doibase 10.1103/PhysRevD.77.013008} {\bibfield
  {journal} {\bibinfo  {journal} {Phys. Rev. D}\ }\textbf {\bibinfo {volume}
  {77}},\ \bibinfo {pages} {013008} (\bibinfo {year} {2008})},\ \Eprint
  {http://arxiv.org/abs/0708.1344} {arXiv:0708.1344 [hep-ph]} \BibitemShut
  {NoStop}%
\bibitem [{\citenamefont {Fridell}\ \emph
  {et~al.}(2024{\natexlab{a}})\citenamefont {Fridell}, \citenamefont {Gr\'af},
  \citenamefont {Harz},\ and\ \citenamefont {Hati}}]{Fridell:2024pmw}%
  \BibitemOpen
  \bibfield  {author} {\bibinfo {author} {\bibfnamefont {K.}~\bibnamefont
  {Fridell}}, \bibinfo {author} {\bibfnamefont {L.}~\bibnamefont {Gr\'af}},
  \bibinfo {author} {\bibfnamefont {J.}~\bibnamefont {Harz}}, \ and\ \bibinfo
  {author} {\bibfnamefont {C.}~\bibnamefont {Hati}},\ }\href@noop {} {\
  (\bibinfo {year} {2024}{\natexlab{a}})},\ \Eprint
  {http://arxiv.org/abs/2412.14268} {arXiv:2412.14268 [hep-ph]} \BibitemShut
  {NoStop}%
\bibitem [{Note1()}]{Note1}%
  \BibitemOpen
  \bibinfo {note} {This naive correlation between contribution to absolute
  light neutrino mass and the light-neutrino-exchange mechanism contribution to
  $0\nu \beta \beta $ (often parametrised by the effective neutrino mass
  parameter $m_{\beta \beta }$ at low-energies) can be broken e.g.~ if a
  cancellation can be arranged such that $m_{\beta \beta }$ is vanishingly
  small but the rate of $0\nu \beta \beta $ is not.}\BibitemShut {Stop}%
\bibitem [{\citenamefont {Cirigliano}\ \emph {et~al.}(2017)\citenamefont
  {Cirigliano}, \citenamefont {Dekens}, \citenamefont {de~Vries}, \citenamefont
  {Graesser},\ and\ \citenamefont {Mereghetti}}]{Cirigliano:2017djv}%
  \BibitemOpen
  \bibfield  {author} {\bibinfo {author} {\bibfnamefont {V.}~\bibnamefont
  {Cirigliano}}, \bibinfo {author} {\bibfnamefont {W.}~\bibnamefont {Dekens}},
  \bibinfo {author} {\bibfnamefont {J.}~\bibnamefont {de~Vries}}, \bibinfo
  {author} {\bibfnamefont {M.~L.}\ \bibnamefont {Graesser}}, \ and\ \bibinfo
  {author} {\bibfnamefont {E.}~\bibnamefont {Mereghetti}},\ }\href {\doibase
  10.1007/JHEP12(2017)082} {\bibfield  {journal} {\bibinfo  {journal} {JHEP}\
  }\textbf {\bibinfo {volume} {12}},\ \bibinfo {pages} {082} (\bibinfo {year}
  {2017})},\ \Eprint {http://arxiv.org/abs/1708.09390} {arXiv:1708.09390
  [hep-ph]} \BibitemShut {NoStop}%
\bibitem [{\citenamefont {Manohar}(2018)}]{Manohar:2018aog}%
  \BibitemOpen
  \bibfield  {author} {\bibinfo {author} {\bibfnamefont {A.~V.}\ \bibnamefont
  {Manohar}},\ }\href {\doibase 10.1093/oso/9780198855743.003.0002} {\
  (\bibinfo {year} {2018}),\ 10.1093/oso/9780198855743.003.0002},\ \Eprint
  {http://arxiv.org/abs/1804.05863} {arXiv:1804.05863 [hep-ph]} \BibitemShut
  {NoStop}%
\bibitem [{\citenamefont {Zhang}(2023)}]{Zhang:2023kvw}%
  \BibitemOpen
  \bibfield  {author} {\bibinfo {author} {\bibfnamefont {D.}~\bibnamefont
  {Zhang}},\ }\href {\doibase 10.1007/JHEP10(2023)148} {\bibfield  {journal}
  {\bibinfo  {journal} {JHEP}\ }\textbf {\bibinfo {volume} {10}},\ \bibinfo
  {pages} {148} (\bibinfo {year} {2023})},\ \Eprint
  {http://arxiv.org/abs/2306.03008} {arXiv:2306.03008 [hep-ph]} \BibitemShut
  {NoStop}%
\bibitem [{\citenamefont {Jenkins}\ \emph {et~al.}(2013)\citenamefont
  {Jenkins}, \citenamefont {Manohar},\ and\ \citenamefont
  {Trott}}]{Jenkins:2013zja}%
  \BibitemOpen
  \bibfield  {author} {\bibinfo {author} {\bibfnamefont {E.~E.}\ \bibnamefont
  {Jenkins}}, \bibinfo {author} {\bibfnamefont {A.~V.}\ \bibnamefont
  {Manohar}}, \ and\ \bibinfo {author} {\bibfnamefont {M.}~\bibnamefont
  {Trott}},\ }\href {\doibase 10.1007/JHEP10(2013)087} {\bibfield  {journal}
  {\bibinfo  {journal} {JHEP}\ }\textbf {\bibinfo {volume} {10}},\ \bibinfo
  {pages} {087} (\bibinfo {year} {2013})},\ \Eprint
  {http://arxiv.org/abs/1308.2627} {arXiv:1308.2627 [hep-ph]} \BibitemShut
  {NoStop}%
\bibitem [{\citenamefont {Jenkins}\ \emph {et~al.}(2014)\citenamefont
  {Jenkins}, \citenamefont {Manohar},\ and\ \citenamefont
  {Trott}}]{Jenkins:2013wua}%
  \BibitemOpen
  \bibfield  {author} {\bibinfo {author} {\bibfnamefont {E.~E.}\ \bibnamefont
  {Jenkins}}, \bibinfo {author} {\bibfnamefont {A.~V.}\ \bibnamefont
  {Manohar}}, \ and\ \bibinfo {author} {\bibfnamefont {M.}~\bibnamefont
  {Trott}},\ }\href {\doibase 10.1007/JHEP01(2014)035} {\bibfield  {journal}
  {\bibinfo  {journal} {JHEP}\ }\textbf {\bibinfo {volume} {01}},\ \bibinfo
  {pages} {035} (\bibinfo {year} {2014})},\ \Eprint
  {http://arxiv.org/abs/1310.4838} {arXiv:1310.4838 [hep-ph]} \BibitemShut
  {NoStop}%
\bibitem [{\citenamefont {Alonso}\ \emph
  {et~al.}(2014{\natexlab{a}})\citenamefont {Alonso}, \citenamefont {Jenkins},
  \citenamefont {Manohar},\ and\ \citenamefont {Trott}}]{Alonso:2013hga}%
  \BibitemOpen
  \bibfield  {author} {\bibinfo {author} {\bibfnamefont {R.}~\bibnamefont
  {Alonso}}, \bibinfo {author} {\bibfnamefont {E.~E.}\ \bibnamefont {Jenkins}},
  \bibinfo {author} {\bibfnamefont {A.~V.}\ \bibnamefont {Manohar}}, \ and\
  \bibinfo {author} {\bibfnamefont {M.}~\bibnamefont {Trott}},\ }\href
  {\doibase 10.1007/JHEP04(2014)159} {\bibfield  {journal} {\bibinfo  {journal}
  {JHEP}\ }\textbf {\bibinfo {volume} {04}},\ \bibinfo {pages} {159} (\bibinfo
  {year} {2014}{\natexlab{a}})},\ \Eprint {http://arxiv.org/abs/1312.2014}
  {arXiv:1312.2014 [hep-ph]} \BibitemShut {NoStop}%
\bibitem [{\citenamefont {Alonso}\ \emph
  {et~al.}(2014{\natexlab{b}})\citenamefont {Alonso}, \citenamefont {Chang},
  \citenamefont {Jenkins}, \citenamefont {Manohar},\ and\ \citenamefont
  {Shotwell}}]{Alonso:2014zka}%
  \BibitemOpen
  \bibfield  {author} {\bibinfo {author} {\bibfnamefont {R.}~\bibnamefont
  {Alonso}}, \bibinfo {author} {\bibfnamefont {H.-M.}\ \bibnamefont {Chang}},
  \bibinfo {author} {\bibfnamefont {E.~E.}\ \bibnamefont {Jenkins}}, \bibinfo
  {author} {\bibfnamefont {A.~V.}\ \bibnamefont {Manohar}}, \ and\ \bibinfo
  {author} {\bibfnamefont {B.}~\bibnamefont {Shotwell}},\ }\href {\doibase
  10.1016/j.physletb.2014.05.065} {\bibfield  {journal} {\bibinfo  {journal}
  {Phys. Lett. B}\ }\textbf {\bibinfo {volume} {734}},\ \bibinfo {pages} {302}
  (\bibinfo {year} {2014}{\natexlab{b}})},\ \Eprint
  {http://arxiv.org/abs/1405.0486} {arXiv:1405.0486 [hep-ph]} \BibitemShut
  {NoStop}%
\bibitem [{\citenamefont {Babu}\ \emph {et~al.}(1993)\citenamefont {Babu},
  \citenamefont {Leung},\ and\ \citenamefont {Pantaleone}}]{Babu:1993qv}%
  \BibitemOpen
  \bibfield  {author} {\bibinfo {author} {\bibfnamefont {K.~S.}\ \bibnamefont
  {Babu}}, \bibinfo {author} {\bibfnamefont {C.~N.}\ \bibnamefont {Leung}}, \
  and\ \bibinfo {author} {\bibfnamefont {J.~T.}\ \bibnamefont {Pantaleone}},\
  }\href {\doibase 10.1016/0370-2693(93)90801-N} {\bibfield  {journal}
  {\bibinfo  {journal} {Phys. Lett. B}\ }\textbf {\bibinfo {volume} {319}},\
  \bibinfo {pages} {191} (\bibinfo {year} {1993})},\ \Eprint
  {http://arxiv.org/abs/hep-ph/9309223} {arXiv:hep-ph/9309223} \BibitemShut
  {NoStop}%
\bibitem [{\citenamefont {Chankowski}\ and\ \citenamefont
  {Pluciennik}(1993)}]{Chankowski:1993tx}%
  \BibitemOpen
  \bibfield  {author} {\bibinfo {author} {\bibfnamefont {P.~H.}\ \bibnamefont
  {Chankowski}}\ and\ \bibinfo {author} {\bibfnamefont {Z.}~\bibnamefont
  {Pluciennik}},\ }\href {\doibase 10.1016/0370-2693(93)90330-K} {\bibfield
  {journal} {\bibinfo  {journal} {Phys. Lett. B}\ }\textbf {\bibinfo {volume}
  {316}},\ \bibinfo {pages} {312} (\bibinfo {year} {1993})},\ \Eprint
  {http://arxiv.org/abs/hep-ph/9306333} {arXiv:hep-ph/9306333} \BibitemShut
  {NoStop}%
\bibitem [{\citenamefont {Antusch}\ \emph {et~al.}(2001)\citenamefont
  {Antusch}, \citenamefont {Drees}, \citenamefont {Kersten}, \citenamefont
  {Lindner},\ and\ \citenamefont {Ratz}}]{Antusch:2001ck}%
  \BibitemOpen
  \bibfield  {author} {\bibinfo {author} {\bibfnamefont {S.}~\bibnamefont
  {Antusch}}, \bibinfo {author} {\bibfnamefont {M.}~\bibnamefont {Drees}},
  \bibinfo {author} {\bibfnamefont {J.}~\bibnamefont {Kersten}}, \bibinfo
  {author} {\bibfnamefont {M.}~\bibnamefont {Lindner}}, \ and\ \bibinfo
  {author} {\bibfnamefont {M.}~\bibnamefont {Ratz}},\ }\href {\doibase
  10.1016/S0370-2693(01)01127-3} {\bibfield  {journal} {\bibinfo  {journal}
  {Phys. Lett. B}\ }\textbf {\bibinfo {volume} {519}},\ \bibinfo {pages} {238}
  (\bibinfo {year} {2001})},\ \Eprint {http://arxiv.org/abs/hep-ph/0108005}
  {arXiv:hep-ph/0108005} \BibitemShut {NoStop}%
\bibitem [{\citenamefont {Chala}\ and\ \citenamefont
  {Titov}(2021)}]{Chala:2021juk}%
  \BibitemOpen
  \bibfield  {author} {\bibinfo {author} {\bibfnamefont {M.}~\bibnamefont
  {Chala}}\ and\ \bibinfo {author} {\bibfnamefont {A.}~\bibnamefont {Titov}},\
  }\href {\doibase 10.1103/PhysRevD.104.035002} {\bibfield  {journal} {\bibinfo
   {journal} {Phys. Rev. D}\ }\textbf {\bibinfo {volume} {104}},\ \bibinfo
  {pages} {035002} (\bibinfo {year} {2021})},\ \Eprint
  {http://arxiv.org/abs/2104.08248} {arXiv:2104.08248 [hep-ph]} \BibitemShut
  {NoStop}%
\bibitem [{\citenamefont {Liao}\ and\ \citenamefont {Ma}(2019)}]{Liao:2019tep}%
  \BibitemOpen
  \bibfield  {author} {\bibinfo {author} {\bibfnamefont {Y.}~\bibnamefont
  {Liao}}\ and\ \bibinfo {author} {\bibfnamefont {X.-D.}\ \bibnamefont {Ma}},\
  }\href {\doibase 10.1007/JHEP03(2019)179} {\bibfield  {journal} {\bibinfo
  {journal} {JHEP}\ }\textbf {\bibinfo {volume} {03}},\ \bibinfo {pages} {179}
  (\bibinfo {year} {2019})},\ \Eprint {http://arxiv.org/abs/1901.10302}
  {arXiv:1901.10302 [hep-ph]} \BibitemShut {NoStop}%
\bibitem [{\citenamefont {Zhang}(2024)}]{Zhang:2023ndw}%
  \BibitemOpen
  \bibfield  {author} {\bibinfo {author} {\bibfnamefont {D.}~\bibnamefont
  {Zhang}},\ }\href {\doibase 10.1007/JHEP02(2024)133} {\bibfield  {journal}
  {\bibinfo  {journal} {JHEP}\ }\textbf {\bibinfo {volume} {02}},\ \bibinfo
  {pages} {133} (\bibinfo {year} {2024})},\ \Eprint
  {http://arxiv.org/abs/2310.11055} {arXiv:2310.11055 [hep-ph]} \BibitemShut
  {NoStop}%
\bibitem [{Note2()}]{Note2}%
  \BibitemOpen
  \bibinfo {note} {In Ref.~\cite {Zhang:2023kvw}, it was shown that $\protect
  \hat {\gamma }^{(5,5)}$ vanishes for the dimension-5 operator, while $\gamma
  ^{(7,5)}$ vanishes for all dimension-7 operators except one, $\protect
  \mathcal {O}^{(7)}_{LH}$.}\BibitemShut {Stop}%
\bibitem [{\citenamefont {Graesser}(2017)}]{Graesser:2016bpz}%
  \BibitemOpen
  \bibfield  {author} {\bibinfo {author} {\bibfnamefont {M.~L.}\ \bibnamefont
  {Graesser}},\ }\href {\doibase 10.1007/JHEP08(2017)099} {\bibfield  {journal}
  {\bibinfo  {journal} {JHEP}\ }\textbf {\bibinfo {volume} {08}},\ \bibinfo
  {pages} {099} (\bibinfo {year} {2017})},\ \Eprint
  {http://arxiv.org/abs/1606.04549} {arXiv:1606.04549 [hep-ph]} \BibitemShut
  {NoStop}%
\bibitem [{\citenamefont {Fridell}\ \emph
  {et~al.}(2024{\natexlab{b}})\citenamefont {Fridell}, \citenamefont {Gr\'af},
  \citenamefont {Harz},\ and\ \citenamefont {Hati}}]{Fridell:2023rtr}%
  \BibitemOpen
  \bibfield  {author} {\bibinfo {author} {\bibfnamefont {K.}~\bibnamefont
  {Fridell}}, \bibinfo {author} {\bibfnamefont {L.}~\bibnamefont {Gr\'af}},
  \bibinfo {author} {\bibfnamefont {J.}~\bibnamefont {Harz}}, \ and\ \bibinfo
  {author} {\bibfnamefont {C.}~\bibnamefont {Hati}},\ }\href {\doibase
  10.1007/JHEP05(2024)154} {\bibfield  {journal} {\bibinfo  {journal} {JHEP}\
  }\textbf {\bibinfo {volume} {05}},\ \bibinfo {pages} {154} (\bibinfo {year}
  {2024}{\natexlab{b}})},\ \Eprint {http://arxiv.org/abs/2306.08709}
  {arXiv:2306.08709 [hep-ph]} \BibitemShut {NoStop}%
\bibitem [{\citenamefont {Scholer}\ \emph {et~al.}(2023)\citenamefont
  {Scholer}, \citenamefont {de~Vries},\ and\ \citenamefont
  {Gr\'af}}]{Scholer:2023bnn}%
  \BibitemOpen
  \bibfield  {author} {\bibinfo {author} {\bibfnamefont {O.}~\bibnamefont
  {Scholer}}, \bibinfo {author} {\bibfnamefont {J.}~\bibnamefont {de~Vries}}, \
  and\ \bibinfo {author} {\bibfnamefont {L.}~\bibnamefont {Gr\'af}},\ }\href
  {\doibase 10.1007/JHEP08(2023)043} {\bibfield  {journal} {\bibinfo  {journal}
  {JHEP}\ }\textbf {\bibinfo {volume} {08}},\ \bibinfo {pages} {043} (\bibinfo
  {year} {2023})},\ \Eprint {http://arxiv.org/abs/2304.05415} {arXiv:2304.05415
  [hep-ph]} \BibitemShut {NoStop}%
\bibitem [{\citenamefont {Cirigliano}\ \emph {et~al.}(2018)\citenamefont
  {Cirigliano}, \citenamefont {Dekens}, \citenamefont {de~Vries}, \citenamefont
  {Graesser},\ and\ \citenamefont {Mereghetti}}]{Cirigliano:2018yza}%
  \BibitemOpen
  \bibfield  {author} {\bibinfo {author} {\bibfnamefont {V.}~\bibnamefont
  {Cirigliano}}, \bibinfo {author} {\bibfnamefont {W.}~\bibnamefont {Dekens}},
  \bibinfo {author} {\bibfnamefont {J.}~\bibnamefont {de~Vries}}, \bibinfo
  {author} {\bibfnamefont {M.~L.}\ \bibnamefont {Graesser}}, \ and\ \bibinfo
  {author} {\bibfnamefont {E.}~\bibnamefont {Mereghetti}},\ }\href {\doibase
  10.1007/JHEP12(2018)097} {\bibfield  {journal} {\bibinfo  {journal} {JHEP}\
  }\textbf {\bibinfo {volume} {12}},\ \bibinfo {pages} {097} (\bibinfo {year}
  {2018})},\ \Eprint {http://arxiv.org/abs/1806.02780} {arXiv:1806.02780
  [hep-ph]} \BibitemShut {NoStop}%
\bibitem [{\citenamefont {Kotila}\ and\ \citenamefont
  {Iachello}(2012)}]{Kotila:2012zza}%
  \BibitemOpen
  \bibfield  {author} {\bibinfo {author} {\bibfnamefont {J.}~\bibnamefont
  {Kotila}}\ and\ \bibinfo {author} {\bibfnamefont {F.}~\bibnamefont
  {Iachello}},\ }\href {\doibase 10.1103/PhysRevC.85.034316} {\bibfield
  {journal} {\bibinfo  {journal} {Phys. Rev. C}\ }\textbf {\bibinfo {volume}
  {85}},\ \bibinfo {pages} {034316} (\bibinfo {year} {2012})},\ \Eprint
  {http://arxiv.org/abs/1209.5722} {arXiv:1209.5722 [nucl-th]} \BibitemShut
  {NoStop}%
\bibitem [{\citenamefont {Stefanik}\ \emph {et~al.}(2015)\citenamefont
  {Stefanik}, \citenamefont {Dvornicky}, \citenamefont {Simkovic},\ and\
  \citenamefont {Vogel}}]{Stefanik:2015twa}%
  \BibitemOpen
  \bibfield  {author} {\bibinfo {author} {\bibfnamefont {D.}~\bibnamefont
  {Stefanik}}, \bibinfo {author} {\bibfnamefont {R.}~\bibnamefont {Dvornicky}},
  \bibinfo {author} {\bibfnamefont {F.}~\bibnamefont {Simkovic}}, \ and\
  \bibinfo {author} {\bibfnamefont {P.}~\bibnamefont {Vogel}},\ }\href
  {\doibase 10.1103/PhysRevC.92.055502} {\bibfield  {journal} {\bibinfo
  {journal} {Phys. Rev. C}\ }\textbf {\bibinfo {volume} {92}},\ \bibinfo
  {pages} {055502} (\bibinfo {year} {2015})},\ \Eprint
  {http://arxiv.org/abs/1506.07145} {arXiv:1506.07145 [hep-ph]} \BibitemShut
  {NoStop}%
\bibitem [{\citenamefont {Deppisch}\ \emph {et~al.}(2020)\citenamefont
  {Deppisch}, \citenamefont {Graf}, \citenamefont {Iachello},\ and\
  \citenamefont {Kotila}}]{Deppisch:2020ztt}%
  \BibitemOpen
  \bibfield  {author} {\bibinfo {author} {\bibfnamefont {F.~F.}\ \bibnamefont
  {Deppisch}}, \bibinfo {author} {\bibfnamefont {L.}~\bibnamefont {Graf}},
  \bibinfo {author} {\bibfnamefont {F.}~\bibnamefont {Iachello}}, \ and\
  \bibinfo {author} {\bibfnamefont {J.}~\bibnamefont {Kotila}},\ }\href
  {\doibase 10.1103/PhysRevD.102.095016} {\bibfield  {journal} {\bibinfo
  {journal} {Phys. Rev. D}\ }\textbf {\bibinfo {volume} {102}},\ \bibinfo
  {pages} {095016} (\bibinfo {year} {2020})},\ \Eprint
  {http://arxiv.org/abs/2009.10119} {arXiv:2009.10119 [hep-ph]} \BibitemShut
  {NoStop}%
\bibitem [{\citenamefont {Gasser}\ and\ \citenamefont
  {Leutwyler}(1984)}]{Gasser:1983yg}%
  \BibitemOpen
  \bibfield  {author} {\bibinfo {author} {\bibfnamefont {J.}~\bibnamefont
  {Gasser}}\ and\ \bibinfo {author} {\bibfnamefont {H.}~\bibnamefont
  {Leutwyler}},\ }\href {\doibase 10.1016/0003-4916(84)90242-2} {\bibfield
  {journal} {\bibinfo  {journal} {Annals Phys.}\ }\textbf {\bibinfo {volume}
  {158}},\ \bibinfo {pages} {142} (\bibinfo {year} {1984})}\BibitemShut
  {NoStop}%
\bibitem [{\citenamefont {Weinberg}(1990)}]{Weinberg:1990rz}%
  \BibitemOpen
  \bibfield  {author} {\bibinfo {author} {\bibfnamefont {S.}~\bibnamefont
  {Weinberg}},\ }\href {\doibase 10.1016/0370-2693(90)90938-3} {\bibfield
  {journal} {\bibinfo  {journal} {Phys. Lett. B}\ }\textbf {\bibinfo {volume}
  {251}},\ \bibinfo {pages} {288} (\bibinfo {year} {1990})}\BibitemShut
  {NoStop}%
\bibitem [{\citenamefont {Weinberg}(1991)}]{Weinberg:1991um}%
  \BibitemOpen
  \bibfield  {author} {\bibinfo {author} {\bibfnamefont {S.}~\bibnamefont
  {Weinberg}},\ }\href {\doibase 10.1016/0550-3213(91)90231-L} {\bibfield
  {journal} {\bibinfo  {journal} {Nucl. Phys. B}\ }\textbf {\bibinfo {volume}
  {363}},\ \bibinfo {pages} {3} (\bibinfo {year} {1991})}\BibitemShut {NoStop}%
\bibitem [{\citenamefont {Scherer}(2003)}]{Scherer:2002tk}%
  \BibitemOpen
  \bibfield  {author} {\bibinfo {author} {\bibfnamefont {S.}~\bibnamefont
  {Scherer}},\ }\href@noop {} {\bibfield  {journal} {\bibinfo  {journal} {Adv.
  Nucl. Phys.}\ }\textbf {\bibinfo {volume} {27}},\ \bibinfo {pages} {277}
  (\bibinfo {year} {2003})},\ \Eprint {http://arxiv.org/abs/hep-ph/0210398}
  {arXiv:hep-ph/0210398} \BibitemShut {NoStop}%
\bibitem [{\citenamefont {Scherer}\ and\ \citenamefont
  {Schindler}(2005)}]{Scherer:2005ri}%
  \BibitemOpen
  \bibfield  {author} {\bibinfo {author} {\bibfnamefont {S.}~\bibnamefont
  {Scherer}}\ and\ \bibinfo {author} {\bibfnamefont {M.~R.}\ \bibnamefont
  {Schindler}},\ }\href@noop {} {\  (\bibinfo {year} {2005})},\ \Eprint
  {http://arxiv.org/abs/hep-ph/0505265} {arXiv:hep-ph/0505265} \BibitemShut
  {NoStop}%
\bibitem [{\citenamefont {Bhattacharya}\ \emph {et~al.}(2016)\citenamefont
  {Bhattacharya}, \citenamefont {Cirigliano}, \citenamefont {Cohen},
  \citenamefont {Gupta}, \citenamefont {Lin},\ and\ \citenamefont
  {Yoon}}]{Bhattacharya:2016zcn}%
  \BibitemOpen
  \bibfield  {author} {\bibinfo {author} {\bibfnamefont {T.}~\bibnamefont
  {Bhattacharya}}, \bibinfo {author} {\bibfnamefont {V.}~\bibnamefont
  {Cirigliano}}, \bibinfo {author} {\bibfnamefont {S.}~\bibnamefont {Cohen}},
  \bibinfo {author} {\bibfnamefont {R.}~\bibnamefont {Gupta}}, \bibinfo
  {author} {\bibfnamefont {H.-W.}\ \bibnamefont {Lin}}, \ and\ \bibinfo
  {author} {\bibfnamefont {B.}~\bibnamefont {Yoon}},\ }\href {\doibase
  10.1103/PhysRevD.94.054508} {\bibfield  {journal} {\bibinfo  {journal} {Phys.
  Rev. D}\ }\textbf {\bibinfo {volume} {94}},\ \bibinfo {pages} {054508}
  (\bibinfo {year} {2016})},\ \Eprint {http://arxiv.org/abs/1606.07049}
  {arXiv:1606.07049 [hep-lat]} \BibitemShut {NoStop}%
\bibitem [{\citenamefont {Nicholson}\ \emph {et~al.}(2018)\citenamefont
  {Nicholson} \emph {et~al.}}]{Nicholson:2018mwc}%
  \BibitemOpen
  \bibfield  {author} {\bibinfo {author} {\bibfnamefont {A.}~\bibnamefont
  {Nicholson}} \emph {et~al.},\ }\href {\doibase
  10.1103/PhysRevLett.121.172501} {\bibfield  {journal} {\bibinfo  {journal}
  {Phys. Rev. Lett.}\ }\textbf {\bibinfo {volume} {121}},\ \bibinfo {pages}
  {172501} (\bibinfo {year} {2018})},\ \Eprint
  {http://arxiv.org/abs/1805.02634} {arXiv:1805.02634 [nucl-th]} \BibitemShut
  {NoStop}%
\bibitem [{\citenamefont {Cirigliano}\ \emph
  {et~al.}(2021{\natexlab{a}})\citenamefont {Cirigliano}, \citenamefont
  {Dekens}, \citenamefont {de~Vries}, \citenamefont {Hoferichter},\ and\
  \citenamefont {Mereghetti}}]{Cirigliano:2020dmx}%
  \BibitemOpen
  \bibfield  {author} {\bibinfo {author} {\bibfnamefont {V.}~\bibnamefont
  {Cirigliano}}, \bibinfo {author} {\bibfnamefont {W.}~\bibnamefont {Dekens}},
  \bibinfo {author} {\bibfnamefont {J.}~\bibnamefont {de~Vries}}, \bibinfo
  {author} {\bibfnamefont {M.}~\bibnamefont {Hoferichter}}, \ and\ \bibinfo
  {author} {\bibfnamefont {E.}~\bibnamefont {Mereghetti}},\ }\href {\doibase
  10.1103/PhysRevLett.126.172002} {\bibfield  {journal} {\bibinfo  {journal}
  {Phys. Rev. Lett.}\ }\textbf {\bibinfo {volume} {126}},\ \bibinfo {pages}
  {172002} (\bibinfo {year} {2021}{\natexlab{a}})},\ \Eprint
  {http://arxiv.org/abs/2012.11602} {arXiv:2012.11602 [nucl-th]} \BibitemShut
  {NoStop}%
\bibitem [{\citenamefont {Cirigliano}\ \emph
  {et~al.}(2021{\natexlab{b}})\citenamefont {Cirigliano}, \citenamefont
  {Dekens}, \citenamefont {de~Vries}, \citenamefont {Hoferichter},\ and\
  \citenamefont {Mereghetti}}]{Cirigliano:2021qko}%
  \BibitemOpen
  \bibfield  {author} {\bibinfo {author} {\bibfnamefont {V.}~\bibnamefont
  {Cirigliano}}, \bibinfo {author} {\bibfnamefont {W.}~\bibnamefont {Dekens}},
  \bibinfo {author} {\bibfnamefont {J.}~\bibnamefont {de~Vries}}, \bibinfo
  {author} {\bibfnamefont {M.}~\bibnamefont {Hoferichter}}, \ and\ \bibinfo
  {author} {\bibfnamefont {E.}~\bibnamefont {Mereghetti}},\ }\href {\doibase
  10.1007/JHEP05(2021)289} {\bibfield  {journal} {\bibinfo  {journal} {JHEP}\
  }\textbf {\bibinfo {volume} {05}},\ \bibinfo {pages} {289} (\bibinfo {year}
  {2021}{\natexlab{b}})},\ \Eprint {http://arxiv.org/abs/2102.03371}
  {arXiv:2102.03371 [nucl-th]} \BibitemShut {NoStop}%
\bibitem [{\citenamefont {Wirth}\ \emph {et~al.}(2021)\citenamefont {Wirth},
  \citenamefont {Yao},\ and\ \citenamefont {Hergert}}]{Wirth:2021pij}%
  \BibitemOpen
  \bibfield  {author} {\bibinfo {author} {\bibfnamefont {R.}~\bibnamefont
  {Wirth}}, \bibinfo {author} {\bibfnamefont {J.~M.}\ \bibnamefont {Yao}}, \
  and\ \bibinfo {author} {\bibfnamefont {H.}~\bibnamefont {Hergert}},\ }\href
  {\doibase 10.1103/PhysRevLett.127.242502} {\bibfield  {journal} {\bibinfo
  {journal} {Phys. Rev. Lett.}\ }\textbf {\bibinfo {volume} {127}},\ \bibinfo
  {pages} {242502} (\bibinfo {year} {2021})},\ \Eprint
  {http://arxiv.org/abs/2105.05415} {arXiv:2105.05415 [nucl-th]} \BibitemShut
  {NoStop}%
\bibitem [{\citenamefont {Abe}\ \emph {et~al.}(2024)\citenamefont {Abe} \emph
  {et~al.}}]{KamLAND-Zen:2024eml}%
  \BibitemOpen
  \bibfield  {author} {\bibinfo {author} {\bibfnamefont {S.}~\bibnamefont
  {Abe}} \emph {et~al.} (\bibinfo {collaboration} {KamLAND-Zen}),\ }\href@noop
  {} {\  (\bibinfo {year} {2024})},\ \Eprint {http://arxiv.org/abs/2406.11438}
  {arXiv:2406.11438 [hep-ex]} \BibitemShut {NoStop}%
\bibitem [{\citenamefont {Adhikari}\ \emph {et~al.}(2022)\citenamefont
  {Adhikari} \emph {et~al.}}]{nEXO:2021ujk}%
  \BibitemOpen
  \bibfield  {author} {\bibinfo {author} {\bibfnamefont {G.}~\bibnamefont
  {Adhikari}} \emph {et~al.} (\bibinfo {collaboration} {nEXO}),\ }\href
  {\doibase 10.1088/1361-6471/ac3631} {\bibfield  {journal} {\bibinfo
  {journal} {J. Phys. G}\ }\textbf {\bibinfo {volume} {49}},\ \bibinfo {pages}
  {015104} (\bibinfo {year} {2022})},\ \Eprint
  {http://arxiv.org/abs/2106.16243} {arXiv:2106.16243 [nucl-ex]} \BibitemShut
  {NoStop}%
\bibitem [{\citenamefont {Abe}\ \emph {et~al.}(2023)\citenamefont {Abe} \emph
  {et~al.}}]{KamLAND-Zen:2022tow}%
  \BibitemOpen
  \bibfield  {author} {\bibinfo {author} {\bibfnamefont {S.}~\bibnamefont
  {Abe}} \emph {et~al.} (\bibinfo {collaboration} {KamLAND-Zen}),\ }\href
  {\doibase 10.1103/PhysRevLett.130.051801} {\bibfield  {journal} {\bibinfo
  {journal} {Phys. Rev. Lett.}\ }\textbf {\bibinfo {volume} {130}},\ \bibinfo
  {pages} {051801} (\bibinfo {year} {2023})},\ \Eprint
  {http://arxiv.org/abs/2203.02139} {arXiv:2203.02139 [hep-ex]} \BibitemShut
  {NoStop}%
\bibitem [{Note3()}]{Note3}%
  \BibitemOpen
  \bibinfo {note} {If the heavy degrees of freedom are hierarchical in mass,
  then one should employ step-by-step matching and mixing integerating out one
  heavy degree of freedom at a time starting from the UV model. In such a
  scenario, one would expect logarithmic corrections due to the running between
  the two hierarchical heavier scales.}\BibitemShut {Stop}%
\bibitem [{Note4()}]{Note4}%
  \BibitemOpen
  \bibinfo {note} {For the figure we make the benchmark choice of fixing the NP
  couplings to unity. However, for smaller values of the NP couplings, both
  dimension-5 and -7 lines shift upwards, with slightly different slopes
  because of a different $\Lambda $ dependence.}\BibitemShut {Stop}%
\bibitem [{Note5()}]{Note5}%
  \BibitemOpen
  \bibinfo {note} {The behaviour of the different contributions and the
  conclusion remains very similar if we instead assume a flavour universal
  structure for $f_1$ Yukawa matrix, with slight shift in scale.}\BibitemShut
  {Stop}%
\bibitem [{\citenamefont {Babu}\ \emph {et~al.}(2009)\citenamefont {Babu},
  \citenamefont {Nandi},\ and\ \citenamefont {Tavartkiladze}}]{Babu:2009aq}%
  \BibitemOpen
  \bibfield  {author} {\bibinfo {author} {\bibfnamefont {K.~S.}\ \bibnamefont
  {Babu}}, \bibinfo {author} {\bibfnamefont {S.}~\bibnamefont {Nandi}}, \ and\
  \bibinfo {author} {\bibfnamefont {Z.}~\bibnamefont {Tavartkiladze}},\ }\href
  {\doibase 10.1103/PhysRevD.80.071702} {\bibfield  {journal} {\bibinfo
  {journal} {Phys. Rev. D}\ }\textbf {\bibinfo {volume} {80}},\ \bibinfo
  {pages} {071702} (\bibinfo {year} {2009})},\ \Eprint
  {http://arxiv.org/abs/0905.2710} {arXiv:0905.2710 [hep-ph]} \BibitemShut
  {NoStop}%
\bibitem [{\citenamefont {Aoki}\ \emph {et~al.}(2020)\citenamefont {Aoki},
  \citenamefont {Enomoto},\ and\ \citenamefont {Kanemura}}]{Aoki:2020til}%
  \BibitemOpen
  \bibfield  {author} {\bibinfo {author} {\bibfnamefont {M.}~\bibnamefont
  {Aoki}}, \bibinfo {author} {\bibfnamefont {K.}~\bibnamefont {Enomoto}}, \
  and\ \bibinfo {author} {\bibfnamefont {S.}~\bibnamefont {Kanemura}},\ }\href
  {\doibase 10.1103/PhysRevD.101.115019} {\bibfield  {journal} {\bibinfo
  {journal} {Phys. Rev. D}\ }\textbf {\bibinfo {volume} {101}},\ \bibinfo
  {pages} {115019} (\bibinfo {year} {2020})},\ \Eprint
  {http://arxiv.org/abs/2002.12265} {arXiv:2002.12265 [hep-ph]} \BibitemShut
  {NoStop}%
\bibitem [{Note6()}]{Note6}%
  \BibitemOpen
  \bibinfo {note} {Note that we have slightly modified some of the interactions
  to make the quantum number assignments consistent with the interactions,
  which seemed to contain typos in the original reference.}\BibitemShut {Stop}%
\bibitem [{Note7()}]{Note7}%
  \BibitemOpen
  \bibinfo {note} {In the absence of any additional symmetry, the tree-level UV
  completions of $\protect \mathcal {O}_{LHD1}^{(7)}$ involve one of the three
  particles leading to tree-level dimension-5 seesaw realizations. Same is also
  true for $\protect \mathcal {O}_{LHD2}^{(7)}$, $\protect \mathcal
  {O}_{LHW}^{(7)}$ and $\protect \mathcal {O}_{LHB}^{(7)}$.}\BibitemShut
  {Stop}%
\bibitem [{Note8()}]{Note8}%
  \BibitemOpen
  \bibinfo {note} {Interestingly, $\protect \mathcal {O}_{LH}^{(7)}$ (which can
  directly contribute to light neutrino masses after the electroweak symmetry
  breaking) provides the most dominant contribution among all the dimension-7
  operators for $\Lambda >4\times 10^{12}$.}\BibitemShut {Stop}%
\bibitem [{\citenamefont {Fuks}\ \emph {et~al.}(2021)\citenamefont {Fuks},
  \citenamefont {Neundorf}, \citenamefont {Peters}, \citenamefont {Ruiz},\ and\
  \citenamefont {Saimpert}}]{Fuks:2020zbm}%
  \BibitemOpen
  \bibfield  {author} {\bibinfo {author} {\bibfnamefont {B.}~\bibnamefont
  {Fuks}}, \bibinfo {author} {\bibfnamefont {J.}~\bibnamefont {Neundorf}},
  \bibinfo {author} {\bibfnamefont {K.}~\bibnamefont {Peters}}, \bibinfo
  {author} {\bibfnamefont {R.}~\bibnamefont {Ruiz}}, \ and\ \bibinfo {author}
  {\bibfnamefont {M.}~\bibnamefont {Saimpert}},\ }\href {\doibase
  10.1103/PhysRevD.103.115014} {\bibfield  {journal} {\bibinfo  {journal}
  {Phys. Rev. D}\ }\textbf {\bibinfo {volume} {103}},\ \bibinfo {pages}
  {115014} (\bibinfo {year} {2021})},\ \Eprint
  {http://arxiv.org/abs/2012.09882} {arXiv:2012.09882 [hep-ph]} \BibitemShut
  {NoStop}%
\bibitem [{\citenamefont {Deppisch}\ and\ \citenamefont
  {Pas}(2007)}]{Deppisch:2006hb}%
  \BibitemOpen
  \bibfield  {author} {\bibinfo {author} {\bibfnamefont {F.}~\bibnamefont
  {Deppisch}}\ and\ \bibinfo {author} {\bibfnamefont {H.}~\bibnamefont {Pas}},\
  }\href {\doibase 10.1103/PhysRevLett.98.232501} {\bibfield  {journal}
  {\bibinfo  {journal} {Phys. Rev. Lett.}\ }\textbf {\bibinfo {volume} {98}},\
  \bibinfo {pages} {232501} (\bibinfo {year} {2007})},\ \Eprint
  {http://arxiv.org/abs/hep-ph/0612165} {arXiv:hep-ph/0612165} \BibitemShut
  {NoStop}%
\bibitem [{\citenamefont {Gr\'af}\ \emph {et~al.}(2022)\citenamefont {Gr\'af},
  \citenamefont {Lindner},\ and\ \citenamefont {Scholer}}]{Graf:2022lhj}%
  \BibitemOpen
  \bibfield  {author} {\bibinfo {author} {\bibfnamefont {L.}~\bibnamefont
  {Gr\'af}}, \bibinfo {author} {\bibfnamefont {M.}~\bibnamefont {Lindner}}, \
  and\ \bibinfo {author} {\bibfnamefont {O.}~\bibnamefont {Scholer}},\ }\href
  {\doibase 10.1103/PhysRevD.106.035022} {\bibfield  {journal} {\bibinfo
  {journal} {Phys. Rev. D}\ }\textbf {\bibinfo {volume} {106}},\ \bibinfo
  {pages} {035022} (\bibinfo {year} {2022})},\ \Eprint
  {http://arxiv.org/abs/2204.10845} {arXiv:2204.10845 [hep-ph]} \BibitemShut
  {NoStop}%
\bibitem [{\citenamefont {Fuentes-Mart\'\i{}n}\ \emph
  {et~al.}(2023)\citenamefont {Fuentes-Mart\'\i{}n}, \citenamefont {K\"onig},
  \citenamefont {Pag\`es}, \citenamefont {Thomsen},\ and\ \citenamefont
  {Wilsch}}]{Fuentes-Martin:2022jrf}%
  \BibitemOpen
  \bibfield  {author} {\bibinfo {author} {\bibfnamefont {J.}~\bibnamefont
  {Fuentes-Mart\'\i{}n}}, \bibinfo {author} {\bibfnamefont {M.}~\bibnamefont
  {K\"onig}}, \bibinfo {author} {\bibfnamefont {J.}~\bibnamefont {Pag\`es}},
  \bibinfo {author} {\bibfnamefont {A.~E.}\ \bibnamefont {Thomsen}}, \ and\
  \bibinfo {author} {\bibfnamefont {F.}~\bibnamefont {Wilsch}},\ }\href
  {\doibase 10.1140/epjc/s10052-023-11726-1} {\bibfield  {journal} {\bibinfo
  {journal} {Eur. Phys. J. C}\ }\textbf {\bibinfo {volume} {83}},\ \bibinfo
  {pages} {662} (\bibinfo {year} {2023})},\ \Eprint
  {http://arxiv.org/abs/2212.04510} {arXiv:2212.04510 [hep-ph]} \BibitemShut
  {NoStop}%
\bibitem [{\citenamefont {Carmona}\ \emph {et~al.}(2022)\citenamefont
  {Carmona}, \citenamefont {Lazopoulos}, \citenamefont {Olgoso},\ and\
  \citenamefont {Santiago}}]{Carmona:2021xtq}%
  \BibitemOpen
  \bibfield  {author} {\bibinfo {author} {\bibfnamefont {A.}~\bibnamefont
  {Carmona}}, \bibinfo {author} {\bibfnamefont {A.}~\bibnamefont {Lazopoulos}},
  \bibinfo {author} {\bibfnamefont {P.}~\bibnamefont {Olgoso}}, \ and\ \bibinfo
  {author} {\bibfnamefont {J.}~\bibnamefont {Santiago}},\ }\href {\doibase
  10.21468/SciPostPhys.12.6.198} {\bibfield  {journal} {\bibinfo  {journal}
  {SciPost Phys.}\ }\textbf {\bibinfo {volume} {12}},\ \bibinfo {pages} {198}
  (\bibinfo {year} {2022})},\ \Eprint {http://arxiv.org/abs/2112.10787}
  {arXiv:2112.10787 [hep-ph]} \BibitemShut {NoStop}%
\end{thebibliography}%

\begin{appendix}   
\onecolumngrid

\section{End matter}
\twocolumngrid

\prlsection{Notes on the LNV dimension-7 SMEFT basis}{.}
In the literature, several bases for dimension-7 LNV SMEFT operators have been proposed and used. In this work, we follow the same basis as in Ref.~\cite{Cirigliano:2017djv}, 
 \begin{align}\label{our_basis}
    \mathbf{\mathcal{O}^{(7)}_{LH}} =& \eps_{ij}\eps_{mn}(L_i^TCL_m )H_j H_n (H^\dagger H)\, , \nonumber  \\  
    \mathbf{\mathcal{O}^{(7)}_{LHD1}}  =& \eps_{ij}\eps_{mn}(L_i^TC (D_\mu L)_j )H_m  (D^\mu H)_n   \, , \nonumber
          \end{align}
      \begin{align}
    \mathbf{\mathcal{O}^{(7)}_{LHD2}}  =& \eps_{im}\eps_{jn}(L_i^TC (D_\mu L)_j )H_m  (D^\mu H)_n   \, , \nonumber \\
    \mathbf{\mathcal{O}^{(7)}_{LHDe}} =& \eps_{ij}\eps_{mn}(L_i^TC \gamma_\mu e )H_j H_m  (D^\mu H)_n   \, , \nonumber\\
 \mathcal{O}^{(7)}_{LHB}  =& \eps_{ij}\eps_{mn}g'(L_i^TC \sigma^{\mu\nu} L_m ) H_j H_n B_{\mu\nu}  \, ,  \nonumber\\
     \mathbf{\mathcal{O}^{(7)}_{LHW}}  =& \eps_{ij} (\eps\tau^I)_{mn} g_2(L_i^TC \sigma^{\mu\nu} L_m ) H_j H_n W^I_{\mu\nu}  \, , \nonumber\\
     \mathbf{\mathcal{O}^{(7)}_{LL\dbar u D1}} =&  \eps_{ij} (\dbar \gamma_\mu u)(L_i^T C (D^\mu L)_j)\, , \nonumber
           \end{align}
      \begin{align}
     \mathcal{O}_{LL \ebar H}^{(7)}  =& \eps_{ij}\eps_{mn}(\ebar L_i)(L_j^TC L_m )H_n\, , \nonumber\\
     \mathbf{\mathcal{O}^{(7)}_{LL Q\dbar H1}}  =&  \eps_{ij}\eps_{mn}(\dbar L_i)(Q_j^TC L_m )H_n\, , \nonumber\\
     \mathbf{\mathcal{O}^{(7)}_{LL Q\dbar  H2}} =&  \eps_{im}\eps_{jn}(\dbar L_i)(Q_j^TC L_m )H_n\, , \nonumber\\
     \mathbf{\mathcal{O}^{(7)}_{LL \Qbar u  H}}  =& \eps_{ij}(\Qbar_m u)(L_m^TC L_i )H_j\, , \nonumber\\
     \mathbf{\mathcal{O}^{(7)}_{L e u \dbar  H}}  =& \eps_{ij}( L_i^T C\gamma_\mu e)(\dbar \gamma^\mu  u )H_j \, ,
     \end{align}
     where the operators in bold contribute to the \0 directly at the tree level. Another useful basis proposed in the literature is the flavour symmetric basis in Refs.~\cite{Zhang:2023kvw,Zhang:2023ndw}, which provides a convenient form for the RGEs. While the convention for some of the operators in this basis is shared with Eq. \eqref{our_basis}, the different ones are related as follows
\begin{align}
    C^{(S) \alpha\beta}_{\ell H} =& \frac{1}{2} \left\{ C^{ \alpha\beta}_{L H} + C^{ \beta\alpha}_{L H}\right\}, \nonumber \\
    C^{(S) \alpha\beta}_{\ell H D 1} =& \frac{1}{2} \left\{ C^{ \alpha\beta}_{L H D 1} + C^{ \beta\alpha}_{L H D 1}\right\}, \nonumber \\
    C^{(S) \alpha\beta}_{\ell H D 2} =& \frac{1}{2} \left\{ C^{ \alpha\beta}_{L H D 2} + C^{ \beta\alpha}_{L H D 2}\right\}, \nonumber \\
    C^{(A)\alpha\beta}_{\ell HB}=& \frac{1}{2 g'} \left\{ C^{\alpha \beta}_{LHB}-C^{\beta \alpha}_{LHB}\right\}, \nonumber\\
    C^{\alpha\beta}_{\ell HW}=& \frac{1}{g_2}C^{\alpha\beta}_{LHW}, \nonumber\\
    C^{\alpha\beta\gamma\lambda}_{\bar{d}u \ell \ell D}=& i C^{\alpha\beta\gamma\lambda}_{LL\bar{d}uD1},\nonumber\\
    C^{\alpha\beta}_{\ell e H D} =& i C^{\alpha\beta}_{LHDe}, \nonumber\\
C^{(S) \alpha\beta}_{\bar{e} \ell \ell \ell H} =& \frac{1}{6} \left( C^{\alpha\beta\gamma\lambda}_{LL\overline{e} H} + C^{\alpha\lambda\beta\gamma}_{LL\overline{e} H} +  C^{\alpha\gamma\lambda\beta}_{LL\overline{e} H} \right.
\nonumber\\
& + \left. C^{\alpha\beta\lambda\gamma}_{LL\overline{e} H} + C^{\alpha\gamma\beta\lambda}_{LL\overline{e} H} +  C^{\alpha\lambda\gamma\beta}_{LL\overline{e} H}
\right),\nonumber\\
    C^{(M) \alpha\beta}_{\bar{e} \ell \ell \ell H} =& \frac{1}{3} \left( C^{\alpha\beta\gamma\lambda}_{LL\overline{e} H} + C^{\alpha\gamma\beta\lambda}_{LL\overline{e} H} - C^{\alpha\lambda\gamma\beta}_{LL\overline{e} H} - C^{\alpha\gamma\lambda\beta}_{LL\overline{e} H} \right), \nonumber \\
    C^{\alpha\beta}_{\bar{d}  \ell q \ell H 1} =& C^{\alpha\beta}_{LLQ\bar{d} H 1},\nonumber\\
    C^{\alpha\beta\gamma\lambda}_{\bar{d}\ell q \ell H2}=& C^{\alpha\beta}_{LLQ\bar{d} H 2}, \nonumber\\
    C^{\alpha\beta}_{\bar{q}  u \ell \ell H}, =& C^{\alpha\beta}_{LL\bar{Q}  u H}, \nonumber \\
    C^{\alpha\beta\gamma\lambda}_{\bar{d} \ell ueH} =& 2 C^{\beta \lambda \alpha \gamma}_{Leu\bar{d}H}.  
\end{align}
In this flavour symmetric basis of Ref.~\cite{Zhang:2023kvw}, the RGE for the evolution of dimension-5 Weinberg operator is given by 
\begin{align} \label{eq:rge-wein5}
	\dot{C}^{\alpha\beta}_5 &= \frac{1}{16 \pi^2} \Bigg[ \left(-3 g_2^2 +  4 \lambda + 2T\right) C_5^{\alpha \beta} \nonumber
    \\&- \frac{3}{2}\left(C^{\alpha \beta}_5 y^\dagger_l Y_l + Y^\dagger_l Y_l T C_5^{\alpha \beta}\right)\nonumber
    \\
    &+ 2 m^2 \left\{ C^{\alpha\beta}_5 \left( 8C^{}_{H\square} - C^{}_{HD} \right) + 8C^{(S)\ast \alpha\beta}_{\ell H}  \right. \nonumber
    \\
     &+ \frac{3}{2} g^2_2 \left( 2 C^{(S)\ast \alpha\beta}_{\ell HD1} + C^{(S)\ast \alpha\beta}_{\ell HD2}  \right)  + \frac{1}{2} \left( Y^{}_l Y^\dagger_l  C^{(S)\dagger}_{\ell HD1} \right)^{\alpha\beta} \nonumber
	\\
	& + \frac{1}{2} \left( Y^{}_l Y^\dagger_l  C^{(S)\dagger}_{\ell HD1} \right)^{\beta\alpha}  - \frac{1}{4} \left( Y^{}_l Y^\dagger_l  C^{(S)\dagger}_{\ell HD2} \right)^{\alpha\beta}  \nonumber
    \\
    &- \frac{1}{4} \left( Y^{}_l Y^\dagger_l  C^{(S)\dagger}_{\ell HD2} \right)^{\beta\alpha} + \left( Y^{}_l C^\dagger_{\ell e HD} \right)^{\alpha\beta}  + \left( Y^{}_l C^\dagger_{\ell e HD}  \right)^{\beta\alpha}\nonumber
	\\
	&- \left( Y^\dagger_l \right)^{}_{\gamma\lambda} \left( 3C^{(S)\ast\gamma\lambda\alpha\beta}_{\overline{e}\ell\ell\ell H}  + C^{(M)\ast\gamma\lambda\alpha\beta}_{\overline{e}\ell\ell\ell H} + C^{(M)\ast\gamma\lambda\beta\alpha}_{\overline{e}\ell\ell\ell H} \right) \nonumber
	\\
	&-    \frac{3}{2} \left( Y^{\dagger}_{\rm d} \right)^{}_{\gamma\lambda} \left( C^{\ast\gamma\alpha\lambda\beta}_{\overline{d}\ell q\ell H1}  + C^{\ast\gamma\beta\lambda\alpha}_{\overline{d}\ell q\ell H1} \right) \nonumber
	\\
    &+ 3 \left.\left( Y^{}_{\rm u} \right)^{}_{\lambda\gamma} \left(  C^{\ast\lambda\gamma\alpha\beta}_{\overline{q}u\ell\ell H}  + C^{\ast\lambda\gamma\beta\alpha}_{\overline{q}u\ell\ell H}  \right)   \right\}\Bigg]  \;,
\end{align}
where 
\begin{equation}
    T=\text{Tr} \left(3Y^\dagger_u Y_u + 3 Y_d^\dagger Y_d + Y^\dagger_l Y_l\right)\, .
\end{equation}
For the complete set of rather lengthy RGE equations for LNV dimension-7 SMEFT operators and their derivations, see, for instance,~\cite {Zhang:2023kvw,Zhang:2023ndw}.
\begin{figure*}[t]
    \centering
    \includegraphics[width=0.9\linewidth]{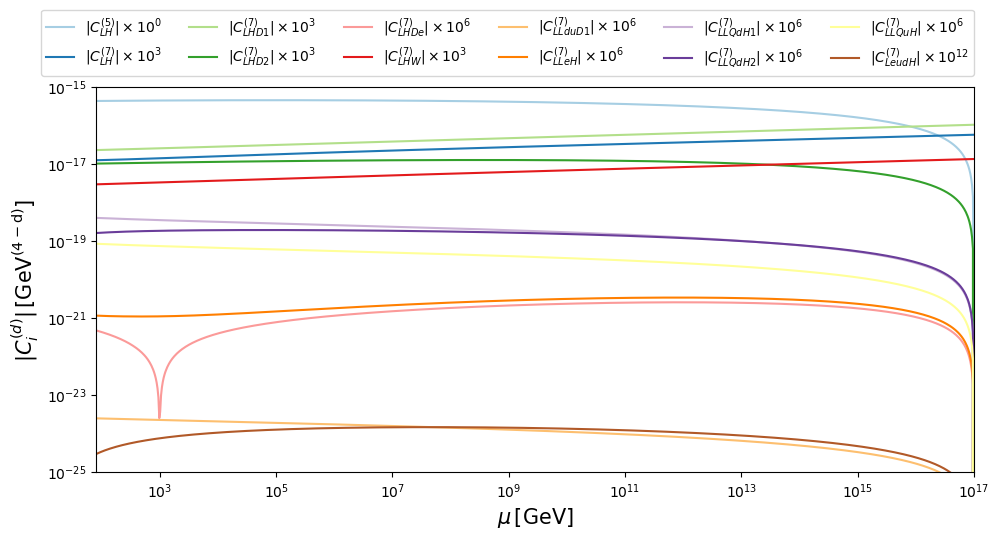}
    \caption{ Plot showing the evolution of various Wilson coefficients as a function of the energy scale $\mu$ for the UV example III in the main text. To obtain this plot, we have assumed one-loop matching at a scale of $10^{17}$ GeV and have included the RGE running effects subsequently until the electroweak symmetry breaking scale. Some of the Wilson coefficients are rescaled for the convenience of presentation, as specified in the legends corresponding to different Wilson coefficients.}
    \label{fig:model3_WC}
\end{figure*}

\prlsection{Notes on Matching of the UV model examples}{.}
The matching and the running of the UV models in the top-down approach must be done using a consistent physical basis or a Green's basis. For some simple UV complete models, only one or a few contributions to the physical basis might be generated. In some cases, however, a large number of Green's basis operators can be generated.  Some tools now provide the possibility to generate the tree-level and one-loop matching contributions for a given UV model in some standard complete Green's basis up to an arbitrary dimension, see e.g. Matchete~\cite{Fuentes-Martin:2022jrf}. Mapping those contributions to a given physical basis can involve tedious matching between redundant and physical operators, applying equations of motion, integration by parts, or algebraic identities to express the generated green basis operators in terms of the complete set of physical basis operators. We have used the packages~\cite{Fuentes-Martin:2022jrf,Carmona:2021xtq} to cross-check some of our loop Matching results. For the sake of completeness. Finally, for UV model example III, the matching contribution to the dimension-5 Weinberg operator is given by
\begin{align}
\Delta C^{\alpha \beta}_5 &= \frac{\kappa y_2^{\alpha} y_1^{\beta} M_{\Sigma'}}{64 \pi^2 (M_{\Sigma'}^2 - M_{\eta}^2)(M_{\Sigma'}^2 - M_{\chi}^2)(M_{\eta}^2 - M_{\chi}^2)} \nonumber\\
&\times \left[  M_{\Sigma'}^2 M_{\eta}^2 \ln \left( \frac{M_{\Sigma'}^2}{M_\eta^2} \right) -\eta\leftrightarrow \chi +\Sigma'\leftrightarrow \chi
\right] \,.
\end{align}
The expressions for the matching contributions to all the other Wilson coefficients induced at the one-loop level are quite lengthy and cannot be included here. However, for the reference of interested readers, we show the evolution of various Wilson coefficients as a function of the energy scale $\mu$ for the UV example III in Fig.~\ref{fig:model3_WC}, assuming one-loop matching at a scale of $ 10^ {17} $ GeV and including the RGE mixing effects subsequently till the electroweak symmetry breaking scale.

\end{appendix}

\end{document}